\documentclass[pra,twocolumn,showpacs,floatfix,superscriptaddress,aps]{revtex4}
\usepackage{amsmath}
\usepackage{makeidx}
\usepackage{amssymb}
\usepackage{graphicx}

\usepackage[usenames]{color}
\usepackage[normalem]{ulem}

\newcommand{\beq}{\begin{equation}}
\newcommand{\eeq}{\end{equation}} 
\newcommand{\beqa}{\begin{eqnarray}}
\newcommand{\eeqa}{\end{eqnarray}} 
\begin{document}

%\begin{multicols}{2}

\title{Demixing and symmetry breaking in  binary dipolar Bose-Einstein condensate solitons}

\author{ S. K. Adhikari\footnote{adhikari@ift.unesp.br; URL: http://www.ift.unesp.br/users/adhikari}
} 
\affiliation{
Instituto de F\'{\i}sica Te\'orica, UNESP - Universidade Estadual Paulista, 01.140-070 S\~ao Paulo, S\~ao Paulo, Brazil
} 

\begin{abstract}

We demonstrate fully   demixed (separated)  
robust and stable bright 
binary dipolar Bose-Einstein condensate soliton in a quasi-one-dimensional   setting
formed due to  
dipolar interactions  for repulsive  
contact interactions. For large  repulsive interspecies 
contact interaction the first species may spatially separate from the second species thus forming 
a demixed configuration, which can be spatially-symmetric or symmetry-broken.
In the spatially-symmetric case, one of the the species occupies the central region, whereas the  other species separates into two equal parts and stay predominantly
out of this central 
region.  In the  symmetry-broken case, the two species stay side by side. 
Stability phase diagrams for the binary solitons are obtained.  
 The results are illustrated with realistic values of parameters in the binary 
$^{164}$Dy-$^{168}$Er and $^{164}$Dy-$^{162}$Dy
mixtures. The demixed solitons are really soliton molecules formed of two types of atoms.  
A proposal for creating dipolar solitons in experiments 
is also presented.

\end{abstract}


\maketitle

\section{Introduction}

Quasi-one-dimensional (quasi-1D) bright
soliton and soliton train were created and investigated 
experimentally in 
Bose-Einstein condensate (BEC) of $^7$Li \cite{1} and $^{85}$Rb atoms 
\cite{3}. These quasi-1D
solitons appear for attractive contact interaction in an  
axially-free BEC under radial harmonic trap \cite{4}.
However, due to collapse instability, in three dimensions (3D),
such solitons are fragile and can accommodate only a small 
number of atoms.

Recent 
observation of  BECs of $^{164}$Dy \cite{ExpDy,dy}, $^{168}$Er \cite{ExpEr} and 
  $^{52}$Cr \cite{cr,saddle} atoms
with large magnetic dipole
moments has opened new directions of research in BEC solitons.
In dipolar BECs, 
in addition to the quasi-1D solitons \cite{1D,1Da}, one could also  have quasi-two-dimensional (quasi-2D) solitons \cite{2D}
and vortex solitons \cite{ol2D}.
 Moreover,  dipolar BEC solitons can be stabilized when the 
harmonic trap(s) is (are) replaced by periodic optical-lattice trap(s)
 in quasi-1D \cite{ol1D} and quasi-2D \cite{ol2D} configurations.
More interestingly, one can have dipolar BEC solitons for fully repulsive 
contact interaction \cite{1D}.  Hence these solitons stabilized solely by long-range dipolar  interaction are less vulnerable to collapse in 3D for a large number of atoms due to the 
repulsive contact interaction. 
 This creates a new 
scenario for robust solitons of large number of atoms stabilized by short-range repulsion and long-range   inter- and intraspecies dipolar attraction. 
One can study the interplay between the long-range
anisotropic dipolar interactions and  variable contact
interactions  using a Feshbach resonance \cite{fesh} in a binary dipolar
BEC soliton.

In a quasi-1D binary dipolar mixture \cite{mfb2,mfb3} free to move along the polarization $z$ direction, for large 
repulsive interspecies contact interaction,
we demonstrate fully demixed (separated) bright dipolar binary solitons
with a minimum of spatial overlap between the two species 
stabilized by 
long-range dipolar interactions
for repulsive  intraspecies contact interactions. 
The demixed binary soliton could be spatially symmetric while 
one of the species occupies the central region ($|z|<z_0$) and  the other species stays symmetrically 
in both peripheral regions ($|z|>z_0$).  For the same set of parameters (number of atoms and dipolar and contact interactions), the binary soliton could also stay in a stable but 
symmetry-broken
spatially 
asymmetric shape where the two species lie side by side along the $z$ direction.

The demixed quasi-1D dipolar binary solitons are really soliton molecules. 
The binding between two (spatial-symmetry-broken)
or three (spatially symmetric)
solitons formed of two types of dipolar atoms is provided by the long-range interspecies dipolar interaction,   which is attractive along the 
axial polarization direction.  Previous consideration of dipolar soliton molecules 
was limited to two   component solitons of the same species of atoms \cite{1D}.

  We consider the numerical solution  
of a three-dimensional mean-field model \cite{mfb2,mfb3}
 of bright binary
dipolar BEC solitons. 
We illustrate our findings using realistic values of inter- and intraspecies 
contact and dipolar interactions 
in the binary dipolar  $^{164}$Dy-$^{168}$Er and  $^{164}$Dy-$^{162}$Dy 
mixtures.  
The repulsive interspecies contact interaction plays a crucial role for 
the appearance of the demixed binary dipolar solitons.
The domain of demixed  solitons is
illustrated in phase diagrams involving critical number of atoms and the interspecies scattering length. The profiles of the binary soliton are displayed in 3D  isodensity plots of the two 
components.    
  The stability of the solitons is demonstrated by considering a  time evolution of the binary soliton when the inter- or intraspecies  contact interaction is perturbed
with a small time-dependent oscillating part. A steady propagation of the binary soliton 
under such a perturbation 
over a long period of time establishes the stability.

As these solitons are stable and robust, they can be created and studied in laboratory and we suggest a way of achieving this goal. 
First,  a highly cigar-shaped dipolar BEC with appropriate number of atoms has to be created. 
Then the axial trap is to be removed slowly and linearly is a small interval of time   while the axially trapped quasi-1D dipolar BEC will turn into an axially free 
quasi-1D dipolar BEC soliton. 
The viability of this procedure is demonstrated by real-time simulation  of the mean-field model.
As the dipolar soliton is an eigenstate of the  mean-field model a very slow  turn-off of the 
axial trap will always lead to a stable soliton.   However, a reasonably quick turn-off  
  is found to lead to a stable soliton as  demonstrated in this paper. 
 A nondipolar soliton is highly fragile and cannot be created in this fashion.

In Sec. II  the time-dependent 3D mean-field model for the binary    dipolar BEC soliton  is presented.  
The results of numerical calculation are shown in Sec. III.  The domain of a stable binary soliton is illustrated in stability phase diagrams showing the maximum  number of atoms  versus
interspecies scattering length for fixed values of inter- and intraspecies
dipolar interactions and intraspecies scattering lengths. The domain of mixed and demixed binary solitons are identified and the demixed 
solitons are further classified by 
spatially-symmetric and symmetry-broken types.  
The 3D isodensity profiles of the symmetric and symmetry-broken binary solitons are contrasted for the same
set of parameters. The viability of preparing these solitons experimentally by relaxing the axial trap on a quasi-1D binary mixture is demonstrated by real-time simulation of the mean-field model. 
A similar demonstration for a single-component dipolar quasi-1D soliton was also made. 
Finally, in Sec. IV   a brief summary of our findings is presented.

\section{Mean-field model for binary mixture}

The extension of the mean-field Gross-Pitaevskii (GP) equation to a binary dipolar
boson-boson  \cite{mfb2,mfb3} and boson-fermion \cite{skabf} mixtures are well established, and, for the sake of completeness, we make a brief summary of the same appropriate for this study.    
We consider a bright binary dipolar BEC soliton, with the 
mass, number of atoms, magnetic  dipole moment, and scattering length for the two species $ i=1,2,$
given by $m_i, N_i, 
\mu_i, a_i,$ respectively.   The intra- ($V_{i}$)
and interspecies ($V_{12}$)
interactions 
for two atoms at  $\bf r$ and $\bf r'$ are 
\begin{eqnarray}\label{intrapot} 
V_i({\bf R})= 
\frac{\mu_0 \mu_i^2}{4\pi}\frac{1-3\cos^2 \theta}{|{\bf R}|^3}+\frac{4\pi 
\hbar^2 a_i}{m_i}\delta({\bf R }),\\
V_{12}({\bf R})= \frac{\mu_0\mu_1  \mu_2}{4\pi}\frac{1-3\cos^2 \theta}{|{\bf R}|^3}+
\frac{2\pi \hbar^2 a_{12}}{m_R}\delta({\bf R}), \label{interpot} 
     \end{eqnarray}
respectively, 
where $a_{12}$ is the intraspecies scattering length, $\mu_0$ is the permeability of free space, 
$\theta$ is the angle made by the vector ${\bf R}$ with the polarization 
$z$ direction,  $\bf R = r-r',$
 and $m_R=m_1m_2/(m_1+m_2)$ is the reduced mass. 
With these interactions, the coupled GP
equations for the binary dipolar BEC can be written as \cite{mfb,mfb2,mfb3}
\begin{align}& \,
{\mbox i} \hbar \frac{\partial \phi_1({\bf r},t)}{\partial t}=
{\Big [}  -\frac{\hbar^2}{2m_1}\nabla^2+
\frac{1}{2}m_1\omega_1^2 \rho^2
\nonumber
\\  & 
+ \frac{4\pi \hbar^2}{m_1}{a}_1 N_1 \vert \phi_1({\bf r},t)\vert^2
+\frac{2\pi \hbar^2}{m_R} {a}_{12} N_2 \vert \phi_2({\bf r},t)|^2
\nonumber \\ &
+ N_1 \frac{ \mu_0 \ { \mu}^2_1 }{4\pi}
\int V_{\mathrm {dd}}({\mathbf R})\vert\phi_1({\mathbf r'},t)\vert^2 d{\mathbf r}'
\nonumber 
\\
 &
+ N_2 \frac{ \mu_0 \ { \mu}_1 \mu_2 }{4\pi}
\int V_{\mathrm {dd}}({\mathbf R})\vert\phi_2({\mathbf r'},t)\vert^2 d{\mathbf r}'
{\Big ]}  \phi_1({\bf r},t),
\label{eq1}
\end{align}
\begin{align}
\label{eq2}
&{\mbox i} \hbar \frac{\partial \phi_2({\bf r},t)}{\partial t}=
{\Big [}  -\frac{\hbar^2}{2m_2}\nabla^2+
\frac{1}{2}m_2\omega_2^2 \rho^2
\nonumber\\ &
+ \frac{4\pi \hbar^2}{m_2}{a}_2 N_2 \vert \phi_2({\bf r},t) \vert^2
+\frac{2\pi \hbar^2}{m_R} {a}_{12} N_1 \vert \phi_1({\bf r},t) \vert^2
\nonumber 
\\ & 
+ N_2 \frac{ \mu_0 \ {\mu}^2_2 }{4\pi}
\int V_{\mathrm {dd}}({\mathbf R})\vert\phi_2({\mathbf r'},t)\vert^2 d{\mathbf r}' 
\nonumber \\ &
+ N_1 \frac{ \mu_0 \ { \mu}_1 \mu_2 }{4\pi}
\int V_{\mathrm {dd}}({\mathbf R})\vert\phi_1({\mathbf r'},t)\vert^2 d{\mathbf r}'
\Big] 
 \phi_2({\bf r},t),
\\&
V_{\mathrm {dd}}({\mathbf R})= 
\frac{1-3\cos^2\theta}{{\mathbf R}^3},  \quad
  \rho^2=x^2+y^2, \quad {\mbox i}=\sqrt{-1}, 
\end{align}
with normalization $\int d{\bf r}| \phi_i({\bf r},t)|^2 =1$.
Here $\omega_i$ are the angular frequencies of the traps acting on the two species in the transverse 
$\rho$ direction. 
The binary soliton is free to move in the axial $z$ direction. 

The intra- and interspecies 
dipolar interactions  are  usually expressed in terms of the dipolar lengths
$a_{\mathrm {dd}}^{(i)}$ and $a_{\mathrm {dd}}^{(12)}$, defined by 
\begin{eqnarray}  a_{\mathrm {dd}}^{(i)}=
\frac{\mu_0  \mu_i^2m_i}{12\pi \hbar ^2    },\quad
a_{\mathrm {dd}}^{(12)}=
\frac{\mu_0  \mu_1 \mu_2m_R}{6\pi \hbar ^2    }.
\end{eqnarray}
We express the strengths of the dipolar 
interactions  by these lengths
and transform Eqs. (\ref{eq1}) and (\ref{eq2}) 
into the following dimensionless form  \cite{mfb2}
\begin{align}& \,
{\mbox i} \frac{\partial \phi_1({\bf r},t)}{\partial t}=
{\Big [}  -\frac{\nabla^2}{2 }
+
\frac{1}{2} \rho^2
\nonumber \\  &  \,
+ g_1 \vert \phi_1 \vert^2
+ g_{12} \vert \phi_2 \vert^2
+ g_{\mathrm {dd}}^{(1)}
\int V_{\mathrm {dd}}({\mathbf R})\vert\phi_1({\mathbf r'},t)
\vert^2 d{\mathbf r}' 
\nonumber \\  &  \,
+ g_{\mathrm {dd}}^{(12)}
\int V_{\mathrm {dd}}({\mathbf R})\vert\phi_2({\mathbf r'},t)
\vert^2 d{\mathbf r}' 
{\Big ]}  \phi_1({\bf r},t),
\label{eq3}
\\
%\end{align}
%\begin{align}
& \,
{\mbox i} \frac{\partial \phi_2({\bf r},t)}{\partial t}={\Big [}  
-m_{12} \frac{\nabla^2}{2}
+
\frac{1}{2}m_\omega \rho^2
\nonumber \\  &  \,
+ g_2 \vert \phi_2 \vert^2 
+ g_{21} \vert \phi_1 \vert^2 
+ g_{\mathrm {dd}}^{(2)}
\int V_{\mathrm {dd}}({\mathbf R})\vert\phi_2({\mathbf r'},t)
\vert^2 d{\mathbf r}'
\nonumber \\ & \,
+ g_{\mathrm {dd}}^{(21)}
\int V_{\mathrm {dd}}({\mathbf R})\vert\phi_1({\mathbf r'},t)
\vert^2 d{\mathbf r}'  
{\Big ]}  \phi_2({\bf r},t),
\label{eq4}
\end{align}
where
%\begin{align}&
 $m_\omega=\omega_2^2/(m_{12}\omega_1^2),$
$m_{12}={m_1}/{m_2},$
$g_1=4\pi a_1 N_1,$
$g_2= 4\pi a_2 N_2 m_{12},$
$g_{12}={2\pi m_1} a_{12} N_2/m_R,$
$g_{21}={2\pi m_1} a_{12} N_1/m_R,$
$g_{\mathrm {dd}}^{(2)}= 3N_2 a_{\mathrm {dd}}^{(2)}m_{12},$
$g_{\mathrm {dd}}^{(1)}= 3N_1 a_{\mathrm {dd}}^{(1)},$
$g_{\mathrm {dd}}^{(12)}= 3N_2 a_{\mathrm {dd}}^{(12)}m_1/(2m_R),$
$g_{\mathrm {dd}}^{(21)}= 3N_1 a_{\mathrm {dd}}^{(12)}m_{1}/(2m_R).$
In Eqs. (\ref{eq3}) and (\ref{eq4}), length is expressed in units of 
oscillator length  $l_0=\sqrt{\hbar/(m_1\omega_1)}$, 
energy in units of oscillator energy  $\hbar\omega_1$, probability density 
$|\phi_i|^2$ in units of $l_0^{-3}$, and time in units of $ 
t_0=1/\omega_1$.

The dimensionless GP equation for a single-component 
dipolar quasi-1D soliton is \cite{1D}
\begin{align}& \,
{\mbox i} \frac{\partial \phi({\bf r},t)}{\partial t}=
{\Big [}  -\frac{\nabla^2}{2 }
+
\frac{1}{2} \rho^2+ 4\pi a N \vert \phi ({\mathbf r},t)\vert^2
\nonumber \\  &  \,
+ 3a_{\mathrm {dd}}N
\int V_{\mathrm {dd}}({\mathbf R})\vert\phi({\mathbf r'},t)
\vert^2 d{\mathbf r}'  
{\Big ]}  \phi({\bf r},t),
\label{single}
\end{align}
where $N$ is the number of atoms, $a$ is the scattering length, $a_{\mathrm {dd}}$ the dipolar length.

\section{Numerical Results}
\label{III}

We demonstrate stable demixed spatially-symmetric and asymmetric  binary dipolar solitons for  realistic values of 
atom numbers and interaction 
parameters 
in the $^{164}$Dy-$^{168}$Er and $^{164}$Dy-$^{162}$Dy
mixtures.
 The $^{164}$Dy and $^{168}$Er atoms 
have the largest magnetic moments of all the dipolar atoms used in BEC experiments. For the $^{164}$Dy-$^{168}$Er mixture,
 the $^{164}$Dy atoms are labeled $i=1$ and  the $^{168}$Er atoms 
are labeled $i=2$.
  The magnetic moment of a $^{164}$Dy  
atom is  $ \mu_1 = 10\mu_B$
\cite{ExpDy} and of a $^{168}$Er atom is $ \mu_2 = 7\mu_B$ \cite{ExpEr}
with 
$\mu_B$ the Bohr magneton leading to the dipolar lengths $a_{\mathrm {dd}}^{(1)}\equiv \mu_0 \mu_1^2 m_1/(12\pi
\hbar^2)\approx 132.7a_0$, 
$a_{\mathrm {dd}}^{(2)}\equiv \mu_0 \mu_2^2 m_2/(12\pi
\hbar^2)\approx 66.6a_0$, and $a_{\mathrm {dd}}^{(12)}\equiv
\mu_0 \mu_1 \mu_2m_R/(6\pi \hbar^2)
\approx 94.0a_0$, with $a_0$ the Bohr radius. 
For the $^{164}$Dy-$^{162}$Dy mixture,  the $^{164}$Dy atoms are labeled $i=1$ and  the $^{162}$Dy atoms 
are labeled $i=2$. The magmetic moment of a $^{162}$Dy atom is 
$ \mu_2 = 10\mu_B$. Consequently the dipolar lengths are:
$a_{\mathrm {dd}}^{(1)}\approx 132.7a_0$,  $a_{\mathrm {dd}}^{(2)}\approx 131.0a_0$, and $a_{\mathrm {dd}}^{(12)}\approx 131.9a_0$.
The fundamental constants used in the evaluation of these dipolar lengths are:
$\mu_B=9.27401 \times 10^{-24}$ Am$^2$,  $a_0 =5.292\times 10^{-11}$ m, $\mu_0=4\pi\times 10^{-7}$ N/A$^2,$ $\hbar= 1.05457\times
10^{-34}$ m$^2$kg/s, 1 amu = $1.66054\times 10^{-27}$ kg.
The dipolar interaction in $^{164}$Dy
atoms is roughly double of that in $^{168}$Er atoms and 
about  eight times larger than that in $^{52}$Cr
atoms with a dipolar length  $a_{\mathrm {dd}}\approx 15a_0$ \cite{cr}.
 We consider the trap frequencies
$\omega_1=\omega_2=2\pi \times 61.6$ Hz, so that the length scale $l_0\equiv
\sqrt{\hbar/m_1\omega_1}\approx 1$ $\mu$m, and  time scale $t_0
\equiv \omega_1^{-1}\approx 2.6$ ms.  
%and the constant $m_\omega= 1/m_{12}$
%in Eq. (\ref{eq4}).

{To the best of our knowledge the experimental values of none of the  scattering lengths considered in this paper are known, except for the intraspecies scattering length of   $^{168}$Er atoms, where preliminary cross-dimensional thermalization
measurements point to a scattering length between $150a_0$ and $200a_0$ \cite{ExpEr}. Nevertheless, scattering lengths can be 
adjusted experimentally using a Feshbach resonance \cite{fesh}.
 }

\begin{figure}[!t]

\begin{center}
\includegraphics[width=\linewidth,clip]{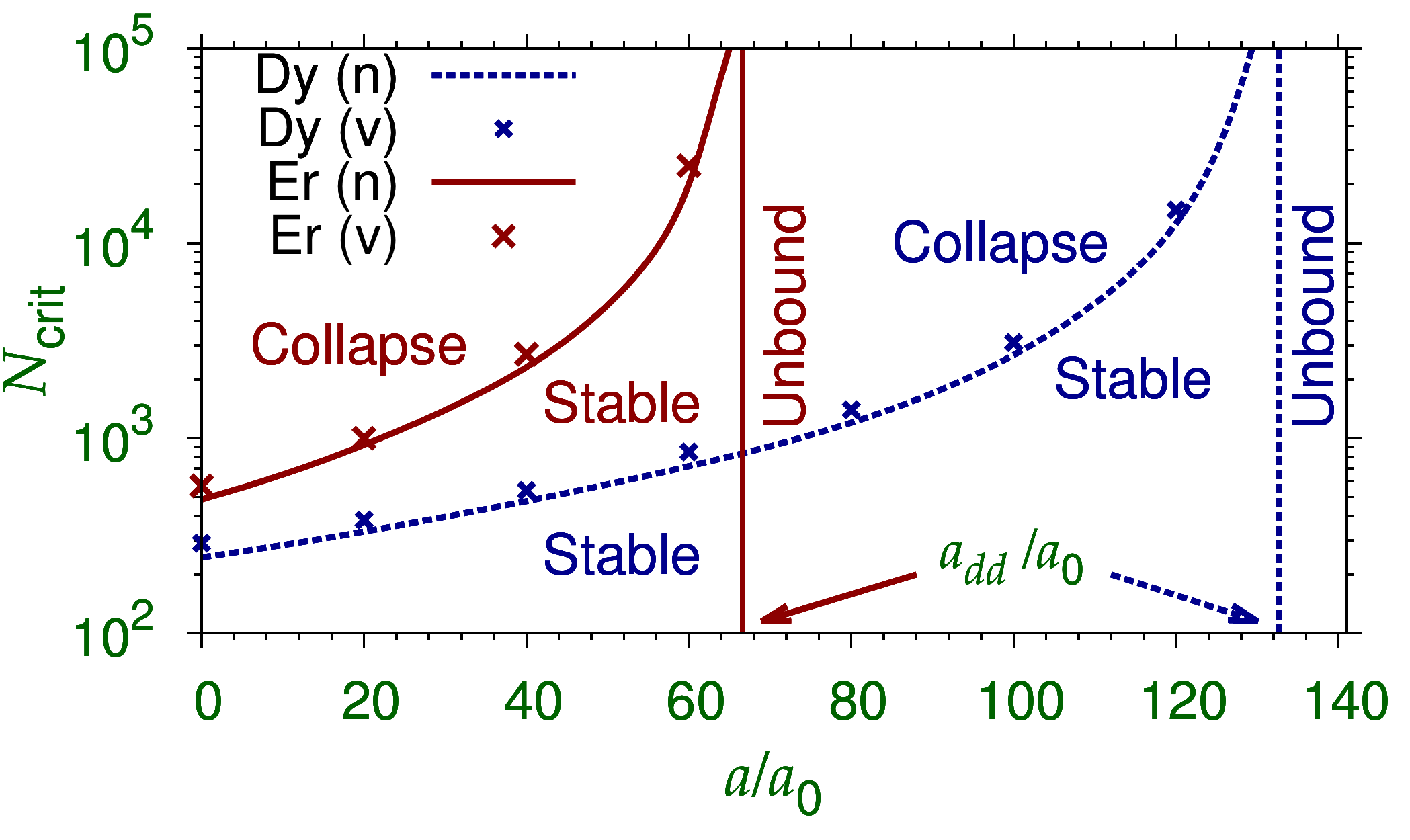}

\caption{ (Color online)   Stability phase diagram illustrating the critical number of atoms 
in a single-component quasi-1D dipolar BEC soliton of $^{164}$Dy or $^{168}$Er atoms from numerical calculation (n) and variational approximation (v). The system is repulsive and unbound for $a\gtrsim a_{\mathrm {dd}}$. Stable quasi-1D solitons appear for $a\lesssim a_{\mathrm {dd}}$ and the number of atoms $N$ below a critical number $N_{\mathrm{crit}}$.
The oscillator length
$l_0 = 1 $ $\mu$m.
}\label{fig1} \end{center}

\end{figure}

We solve the 3D Eqs. (\ref{eq3}) and (\ref{eq4}) 
by the split-step 
Crank-Nicolson discretization scheme using both real- and imaginary-time propagation
  in 3D Cartesian coordinates independent of the underlying 
trap symmetry 
using a space step of 0.1 $\sim$ 0.2
and a time step of 0.001 $\sim$ 0.005 \cite{CPC}.  The dipolar potential term is treated by Fourier transformation  in momentum space using a convolution theorem in usual fashion \cite{Santos01}.

We  study 
demixing in  binary solitons supported by  long-range dipolar interactions
for repulsive interspecies and intraspecies contact interactions.  The repulsive contact interactions make the dipolar solitons less vulnerable to collapse for a larger number of atoms
compared to the nondipolar solitons. 
However, to have a net attraction in the system, atoms with large magnetic dipole moments, such as $^{164}$Dy and $^{168}$Er  are  considered.  In the quasi-1D shape, with radial harmonic trap,
the dipolar interaction is attractive in the axial $z$ direction. The net repulsion in the transverse direction is balanced by the external harmonic trap.

\subsection{Single component $^{164}$Dy and $^{168}$Er solitons}

To have a feeling about the maximum number of atoms in a single-component quasi-1D dipolar soliton, first we solve Eq. (\ref{single}) for different values of the scattering length $a$.
We find that for interaction parameters of $^{164}$Dy and $^{168}$Er atoms the solitons are stable up to a critical maximum number of atoms, beyond which the system collapses. In Fig. \ref{fig1} we plot this critical 
number $N_{\mathrm{crit}}$ versus $a$ from numerical simulation using Eq. (\ref{single}) and  from a Gaussian variational approximation to it as developed in Ref. \cite{1D}.    The variational approximation leads to overbinding and hence can accommodate a larger number of atoms in a stable soliton as can be seen in Fig. \ref{fig1}. 
We find that a stable soliton is possible for
 $a \lesssim a_{\mathrm {dd}}$ and for a number of atoms below this critical number. The critical number of atoms 
increases with the increase of contact repulsion as 
$a\to a_{\mathrm {dd}}$, which is counterintuitive. The solitons are bound by long-range dipolar interaction and increase of contact repulsion gives more stability against collapse for a fixed dipolar interaction strength. 
In this phase diagram three regions are shown: stable, collapse and unbound. In the unbound region ($a\gtrsim a_{\mathrm {dd}}$) contact repulsion dominates over dipolar attraction and the soliton cannot be bound. In the collapse region, the opposite happens and the soliton collapses 
due to excess of attractive dipolar interaction along the axial $z$ direction. In the stable region there is a balance between attraction and repulsion and a stable soliton can be formed.
So in our study of binary solitons we shall take $a \lesssim    a_{\mathrm {dd}}.$   In this fashion, a binary soliton of large number of atoms could be created and this 
would be of greater experimental interest. 
In this  study we take for the $^{164}$Dy atoms $a_1=120a_0$, and for the $^{168}$Er atoms $a_2=60a_0$. From Fig. \ref{fig1} we see that for these values of scattering length a stable bright soliton can accommodate a large number of atoms of the two species. 
 
\begin{figure}[!t]

\begin{center}
\includegraphics[width=\linewidth]{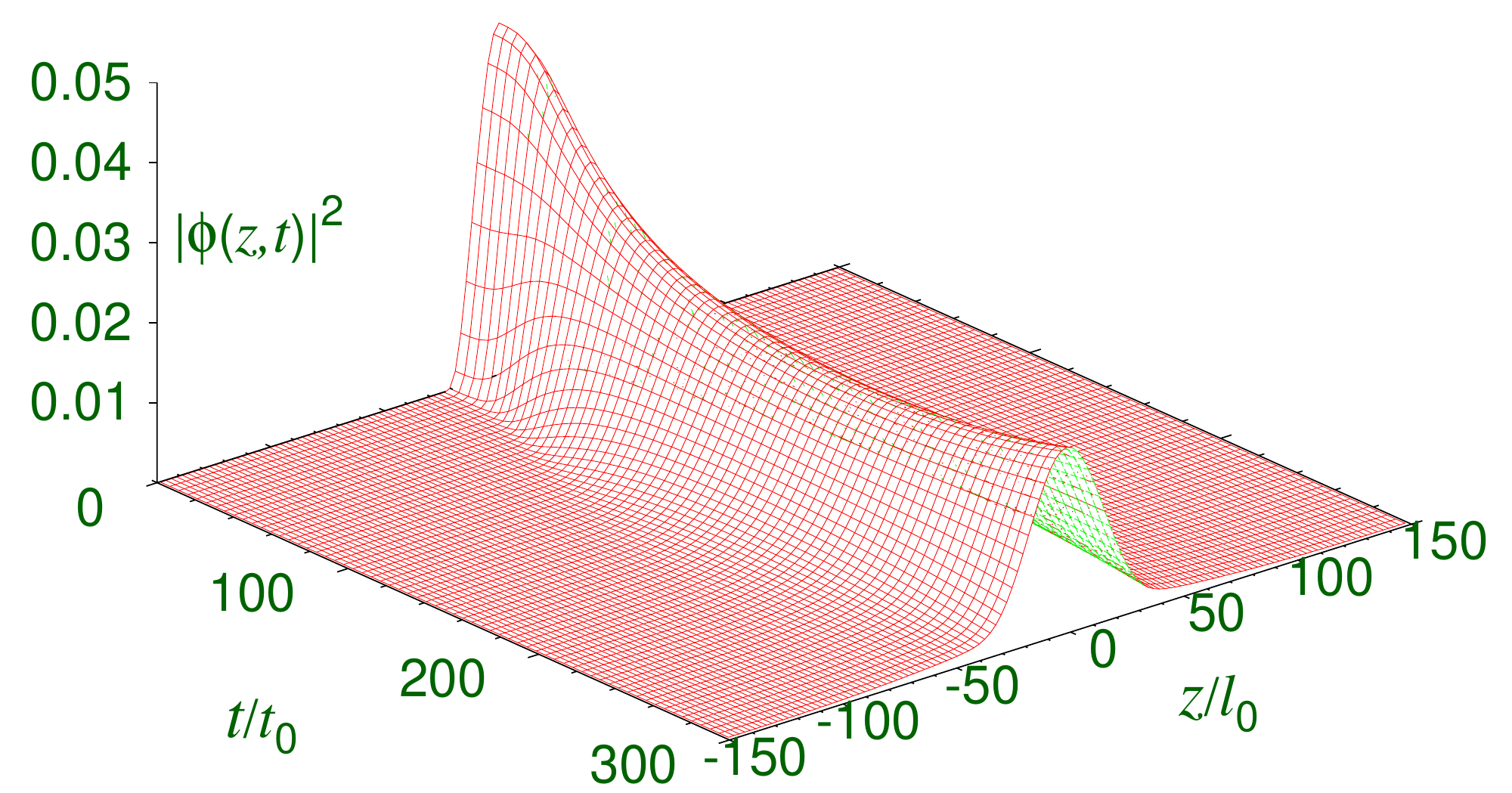}

\caption{ (Color online)
Integrated 1D density $|\phi(z,t)|^2=\int dx\int dy |\phi({\bf r},t)|^2 $ of $^{164}$Dy  atoms during real-time propagation
 when the axial trap of angular frequency $\omega = 2 \pi \times 2$ Hz 
on a quasi-1D dipolar BEC of 8000 $^{164}$Dy atoms is removed linearly for $20>t/t_0>0$ and the resultant 
axially-free soliton is propagated.  
Parameters used in simulation: $a_{\mathrm {dd}}=132.7 a_0, a=120a_0, \omega_\rho=2\pi \times 61$ Hz, $l_0=1$ $\mu$m. 
}\label{fig2} \end{center}

\end{figure}

As the dipolar solitons are stable and robust,  it is 
relatively easy to make these  solitons of a large number of atoms. The nondipolar solitons are usually tiny and fragile containing a small number of atoms \cite{1,3}.  We illustrate a method for creating a dipolar soliton of 8000 $^{164}$Dy atoms by real-time simulation of the GP equation (\ref{single})  with the scattering length $a=120a_0$. By imaginary-time propagation we create a bound quasi-1D BEC 
  in the trap
$V({\bf r})=(\rho^2+ \lambda^2 z^2)/2$ with $\lambda = 0.03$. This corresponds to taking a weak 
axial trap along $z$ axis of angular frequency $\omega_z \approx 2\pi \times 2$ Hz compared to the strong radial trap in the $x-y$ plane of angular frequency  $\omega_\rho=2\pi \times 61$ Hz.    The 3D profile of the BEC as obtained in imaginary-time propagation is used as the initial state in the real-time routine. During real-time propagation,
from $t/t_0=0 $ to 20 the axial trap $\lambda^2 z^2/2$
is gradually (linearly) reduced to zero, so that for $t/t_0>20$ the axially free quasi-1D soliton condition is realized. We continue the real-time propagation for  $t/t_0>20$ for a reasonably large interval of time. Long sustained propagation of the  central peak establishes the soliton nature of the axially-free dipolar BEC. The result of this simulation is presented in Fig. \ref{fig2}, where we  show the integrated 1D density $|\phi(z,t)|^2= \int dx \int dy |\phi({\bf r},t)|^2$ during real-time propagation. Upon
the relaxation of the axial trap reasonably quickly % (in an interval of about 13 ms) 
  the 
soliton expands a little in the beginning before attaining the final shape, 
as can be seen in Fig. \ref{fig2}.   We note that 
the solitonic nature of the axially free dipolar BEC is evident in Fig. \ref{fig2}.  This approach can be used in a laboratory to create dipolar solitons.  Fragile nondipolar solitons 
 supported by contact attraction only cannot be easily  prepared in this fashion.

\subsection{Binary $^{164}$Dy-$^{168}$Er soliton}

Scattering lengths play important roles in the preparation  of binary solitons and 
can be experimentally controlled independently by magnetic \cite{fesh} and optical \cite{opfesh}
Feshbach resonance techniques.
For 
 $^{164}$Dy atoms we take $a_1=120a_0$, and for  $^{168}$Er atoms we take $a_2=60a_0$.
However, the interspecies scattering length $a_{12}$ plays a crucial role in demixing 
 in binary  solitons and will be  considered as a variable.  
 After some experimentation in numerical simulation we realized that a reasonably large interspecies scattering length 
$a_{12}$ ($\gtrsim 70a_0$)  and the number of atoms of the species below a critical value is needed 
for the creation of a demixed binary dipolar soliton. For much larger values of the interspecies scattering length ($a_{12}\gtrsim 120a_0$), the repulsion could be sufficiently strong, so that  solitons of the two species could not be bound to form a 
stable binary dipolar soliton. 
 For larger values of the number of atoms, the system collapses due to a strong 
net dipolar interaction even for repulsive intraspecies interaction. The demixing takes place due 
to interspecies contact repulsion. 
For smaller values of interspecies scattering length, the repulsion  is not sufficient for demixing and a mixed or overlapping 
configuration of binary dipolar BEC soliton is realized \cite{mfb3}.  

A spatially-symmetric binary soliton is obtained by solving Eqs. (\ref{eq3}) and (\ref{eq4}) with imaginary-time propagation with the following spatially-symmetric Gaussian input for both the functions $\phi_i$:
\begin{eqnarray}\label{eq5}
\phi_i({\bf r})= \frac{\pi^{-3/4}}{w_\rho\sqrt w_z} \exp\left[  -\frac{\rho^2}{2w_\rho^2}     -\frac{z^2}{2w_z^2}   \right], 
\end{eqnarray}
where $w$'s are the respective widths.
An initial guess of widths close to their final converged values facilitates convergence.   The input (\ref{eq5})
and the final 
binary soliton are spatially-symmetric around $z=0$.   
To obtain a symmetry-broken binary soliton, a symmetry-broken initial input is needed.

\begin{figure}[!t]

\begin{center}
\includegraphics[width=.8\linewidth]{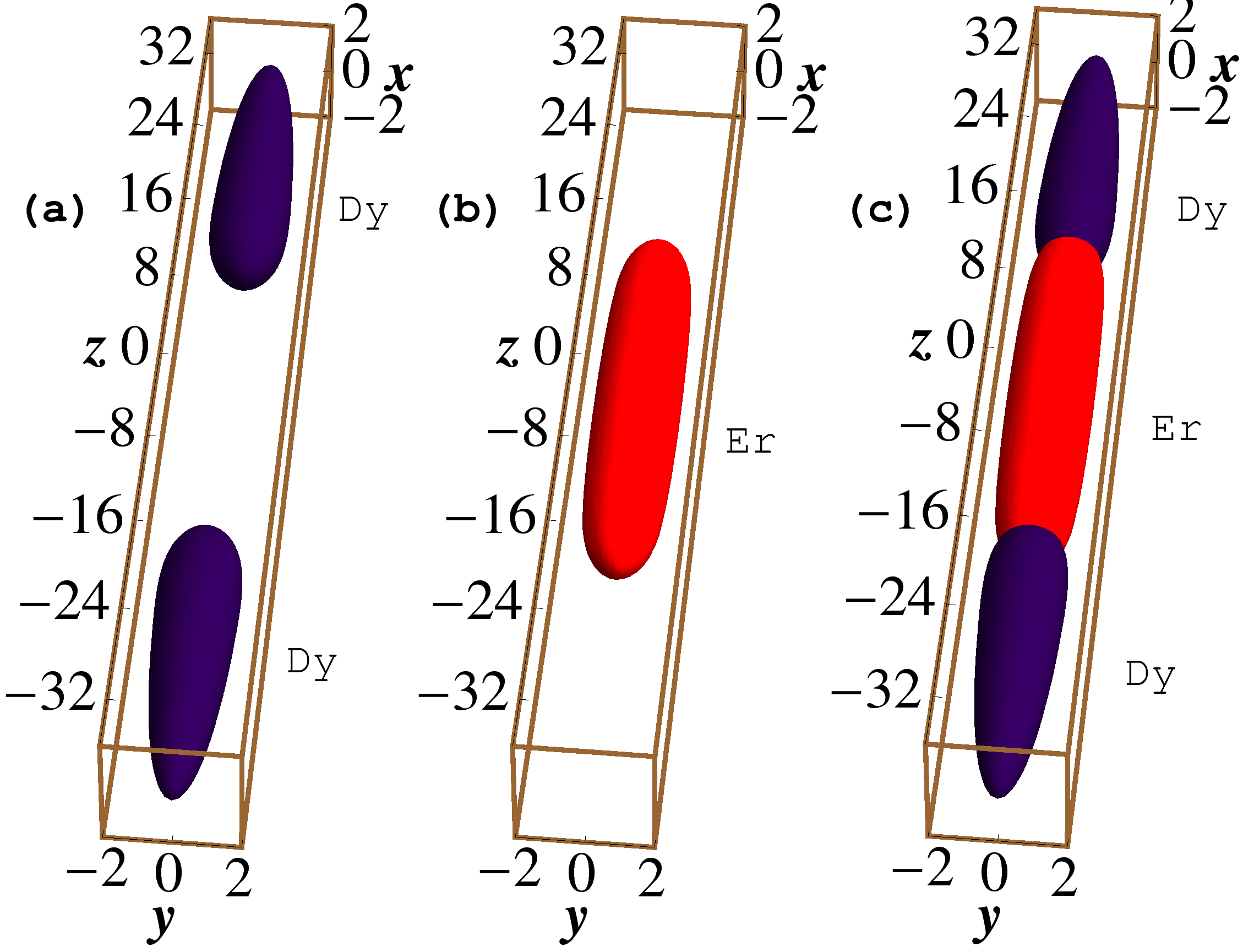}

\caption{ (Color online) 3D isodensity contour 
(a)  $|\phi_1|^2$ of $^{164}$Dy, 
(b)  $ |\phi_2|^2$ of $^{168}$Er, and (c) $(|\phi_1|^2+|\phi_2|^2)$ of the  $^{164}$Dy-$^{168}$Er mixture   for the
binary   soliton of 2000 $^{164}$Dy and 
5000 $^{168}$Er atoms with  $a($Dy$)= 120a_0, a($Er$)= 60a_0, a($Dy-Er$)=90a_0$. The dimensionless lengths $x,y,$ and $z$ are in units of $l_0 (\equiv 1$ $\mu$m).
The density on contour is 0.002.
}\label{fig3} \end{center}

\end{figure}

Now we display the typical profile of a spatially-symmetric demixed  binary dipolar BEC soliton in 
Fig. \ref{fig3} for 2000 $^{164}$Dy atoms and 5000 $^{168}$Er atoms for the interspecies scattering length $a_{12}=a($Dy-Er$)= 90 a_0$ and for large intraspecies scattering lengths: $a_1=a$(Dy) $ = 120a_0$, $a_2=a$(Er) = 60$a_0$. The species containing the larger number of atoms ($^{168}$Er) stay at the center and the species containing the smaller number of atoms ($^{164}$Dy) breaks into two equal parts, leave the central region occupied by $^{168}$Er atoms and stays symmetrically on two sides of the $^{168}$Er soliton with a minimum of interspecies overlap.   Note that the ``3D densities" plotted in this paper are the norms of the respective wave functions $|\phi_i|^2$, whereas the atom densities are   $N_i|\phi_i|^2$.

\begin{figure}[!t]

\begin{center}
\includegraphics[width=\linewidth]{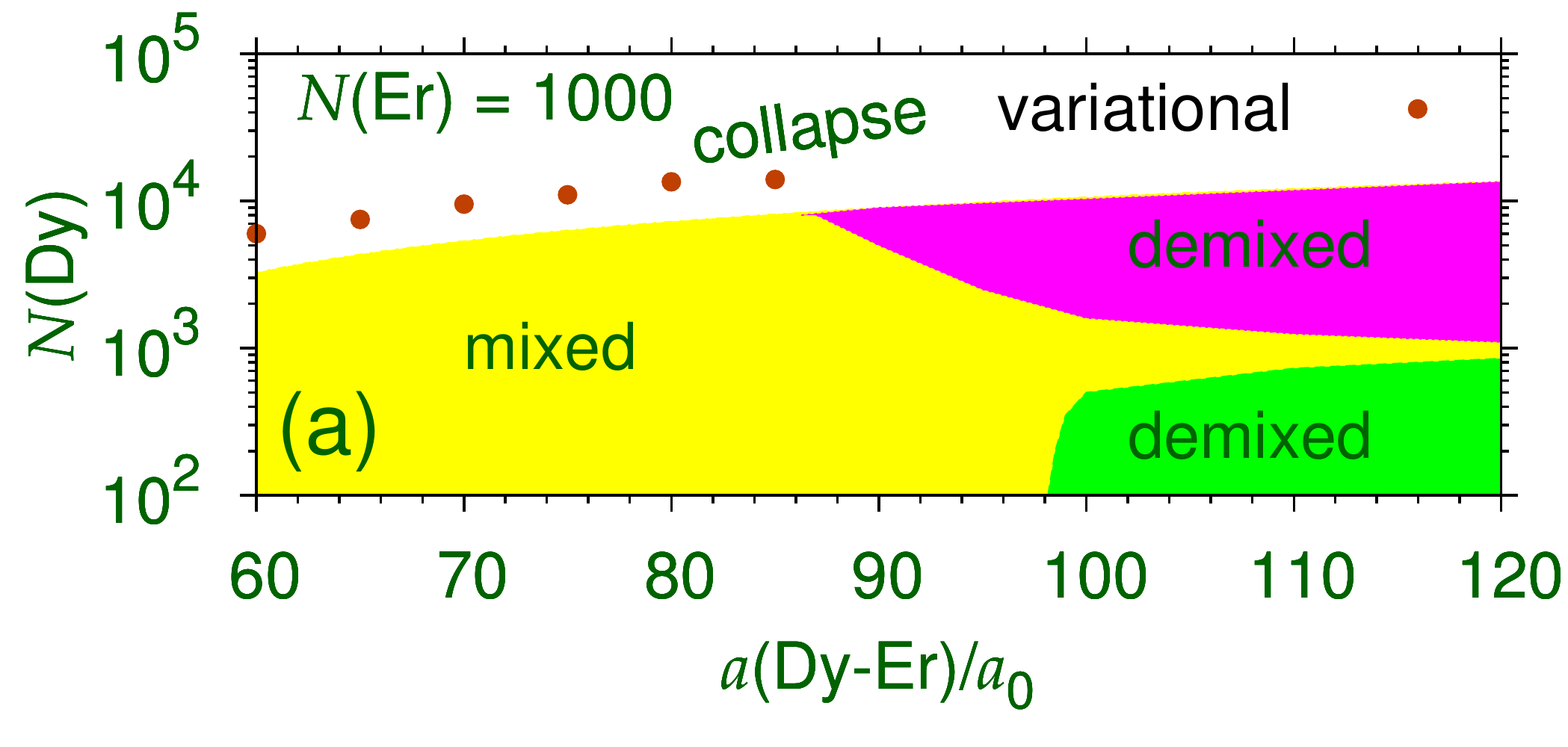}
 \includegraphics[width=\linewidth]{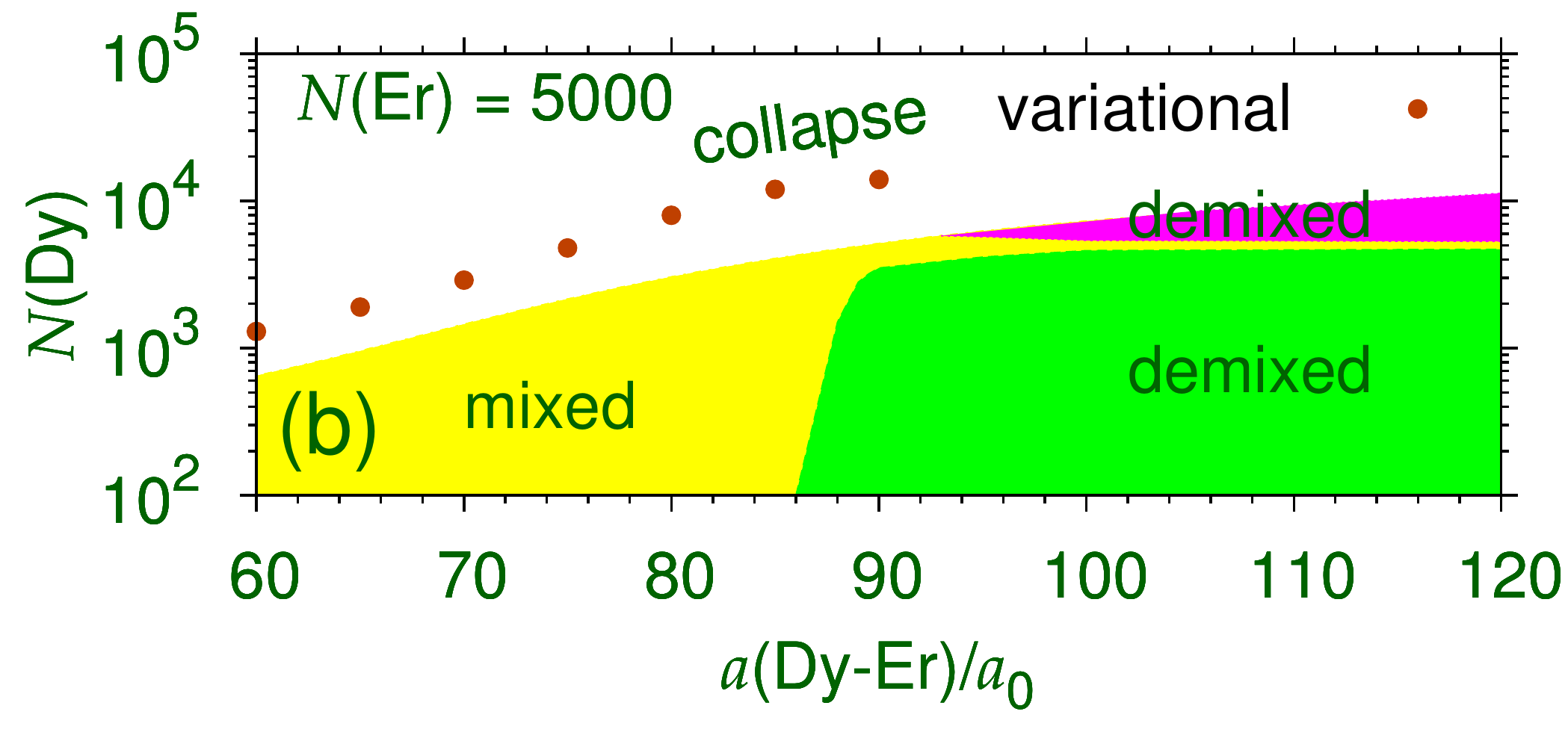}

\caption{ (Color online)  Stability phase diagram for a binary
$^{164}$Dy-$^{168}$Er soliton   for  $N$(Er) = (a) 1000 and (b) 5000 atoms. The variational mixed-collapse border is shown by solid circles. 
The oscillator length $l_0=1$ $\mu$m,   $a$(Dy)$=120a_0,  
a$(Er)$=60a_0.$
Stable demixed binary soliton is observed in the darker (pink and green) region to the right
while stable mixed binary soliton appears in the lighter (yellow) region to the left. For a spatially-symmetric demixed binary soliton,
in the upper darker (pink) region the  $^{168}$Er atoms stay out and in the lower darker (green)
region the $^{164}$Dy atoms stay out of the central region ($z=0$). 
}\label{fig4} \end{center}

\end{figure}

\begin{figure}[!t]

\begin{center}

\includegraphics[width=\linewidth]{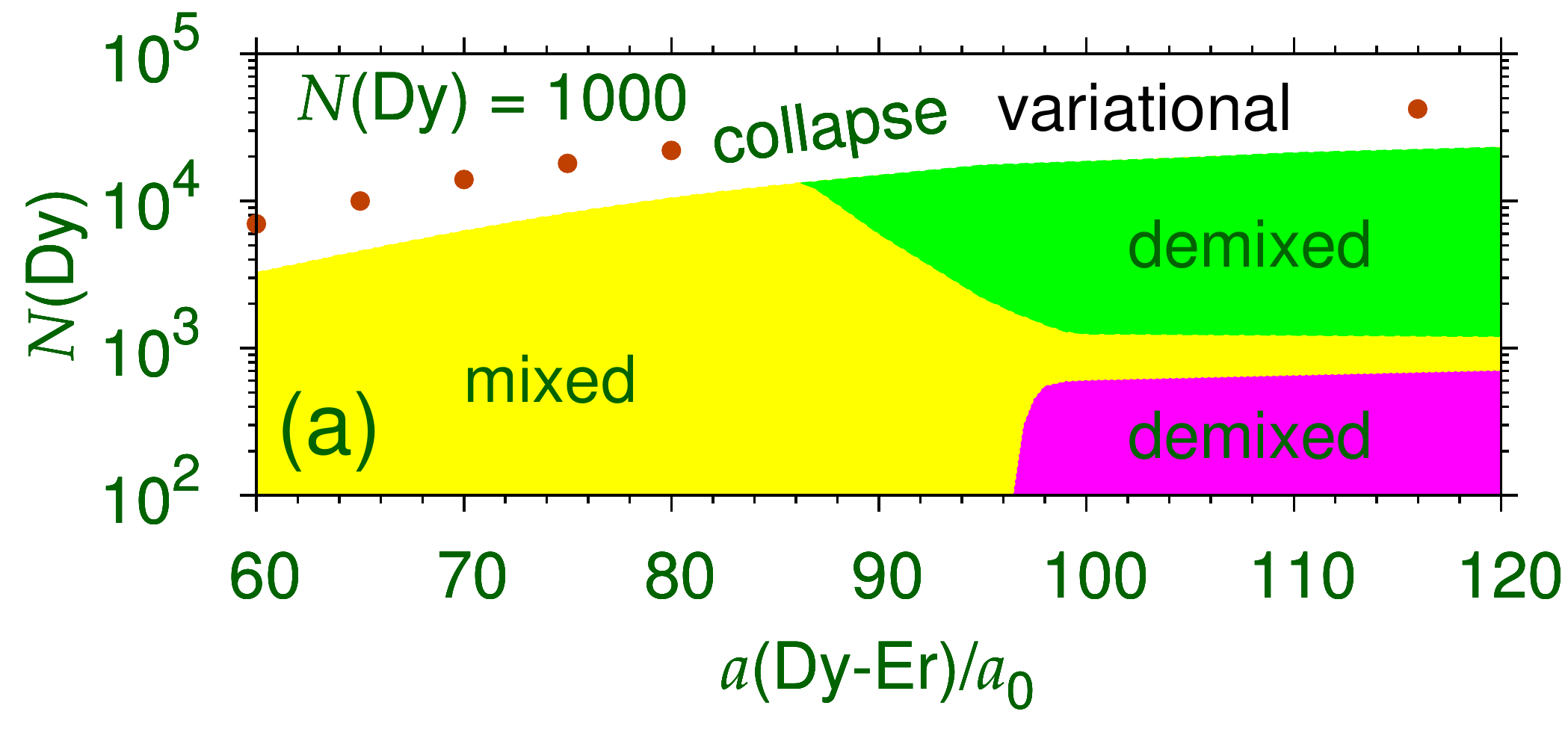}
 \includegraphics[width=\linewidth]{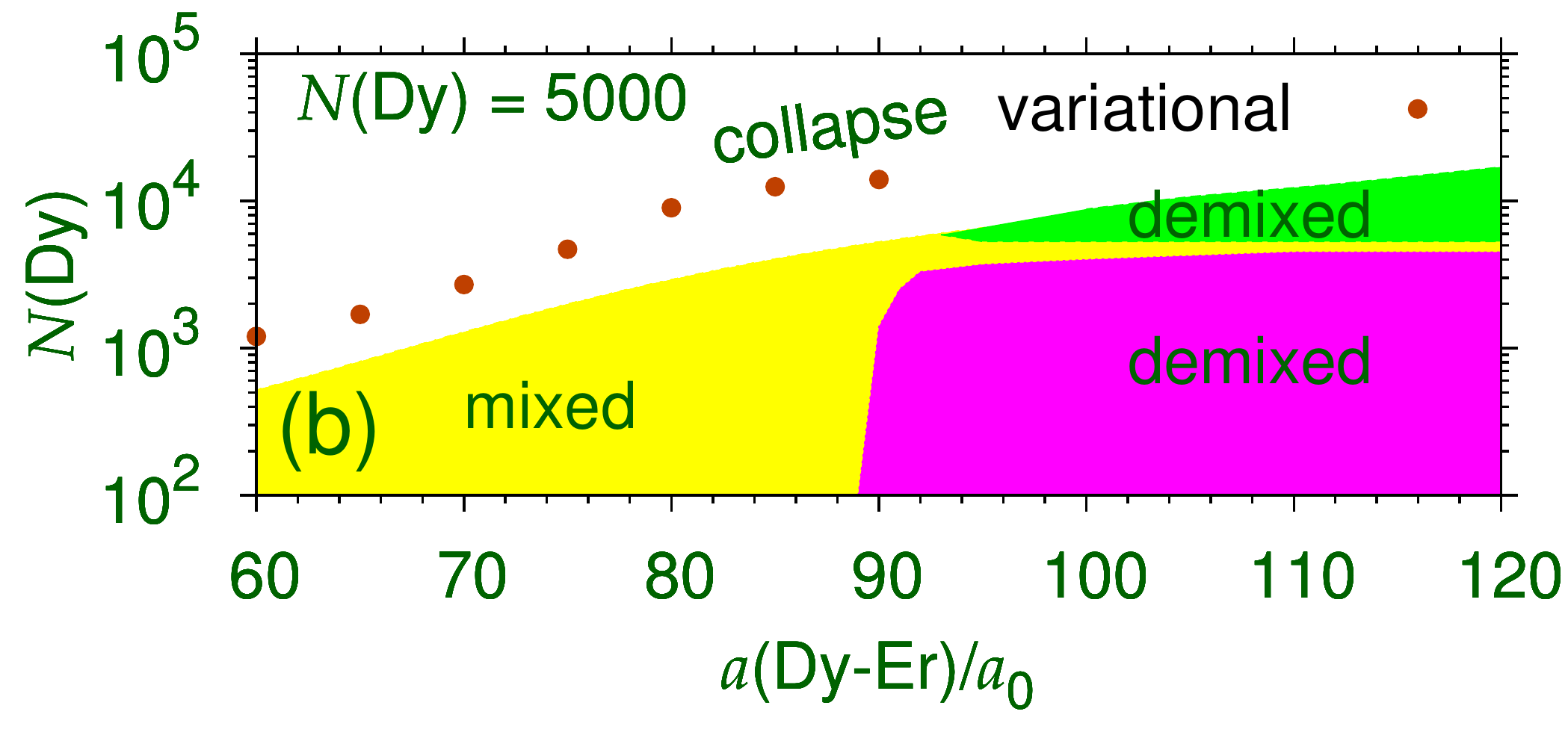}

\caption{ (Color online)
 Same as in Fig. \ref{fig4} for (a) $N$(Dy) = 1000 and (b) 5000 atoms. 
}\label{fig5} \end{center}

\end{figure}

Next we show the stability phase diagrams for the appearance of the demixed binary dipolar 
soliton in the  $^{164}$Dy-$^{168}$Er mixture for different number of atoms and variable 
interspecies scattering length $a_{12}$.  In Figs. \ref{fig4} (a) and (b)  we show the critical number of 
$^{164}$Dy atoms $N_1=N$(Dy) versus  $a_{12}$ for the numbers of $^{168}$Er atoms 
$N_2=N$(Er) $= 1000, 5000$, respectively.
The binary system collapses for $N$(Dy) larger than these critical values. For small values of  $a_{12}$ the binary soliton is mixed and it gets demixed for large  values of  $a_{12}$ due to strong interspecies contact repulsion. 
 A Gaussian variational approximation to Eqs. (\ref{eq3}) and (\ref{eq4}) was developed in Ref. \cite{mfb3} for a mixed binary soliton. This approximation using a mixed configuration of the two solitons can be used to calculate the mixed-collapse boundary. The variational results for this boundary are also shown in    Figs. \ref{fig4}. As the variational approximation overbinds the binary soliton, it shows a larger stable region than the full numerical solution.  
 For $a_{12} \lesssim  80a_0$ the binary solitons are  spatially-symmetric and mixed and the two species of atoms lie on top of each other.  However, for larger $a_{12}$ the contact repulsion between the two species of atoms increases and the system gets demixed.  
 There are two ways this demixing takes place. The demixed binary soliton can either be spatially-symmetric around $z=0$, or it can be spatially asymmetric around $z=0$. 
The two configurations are possible for identical parameters 
of the binary system, e.g., number of atoms of the two components, and all dipolar and contact interaction strengths.  
However, there is a region around 
$N_1\approx N_2$  for these larger values of $a_{12}$, where the solitons continue in a mixed configuration. On both sides of this region demixing can take place.
In the spatially-symmetric configuration, one of the components  (the one containing the larger number of atoms)  continues in the spatially-symmetric state at $z=0.$ The other component 
(the one containing the smaller number of atoms) divides into two equal pieces, separates from each other, leaves the central region occupied by the first component around $z=0$, moves in opposite directions and stays for $|z|>z_0$ in a spatially-symmetric configuration as shown in Fig. \ref{fig3}. In the spatial-symmetry-broken configuration, both components  remain as single pieces and move away from each other and eventually lie on both sides of $z=0$ in a spatially-asymmetric configuration.       In Figs. \ref{fig5} (a) and (b) we show the critical number of 
$^{168}$Er atoms $N_2=N$(Er) versus  $a_{12}$ for the numbers of $^{164}$Dy atoms 
$N_1=N$(Dy) $= 1000, 5000$, respectively.  A similar scenario emerges  in Figs.   \ref{fig5} as 
in Figs. \ref{fig4}.

\begin{figure}[!t]

\begin{center}
\includegraphics[width=.8\linewidth]{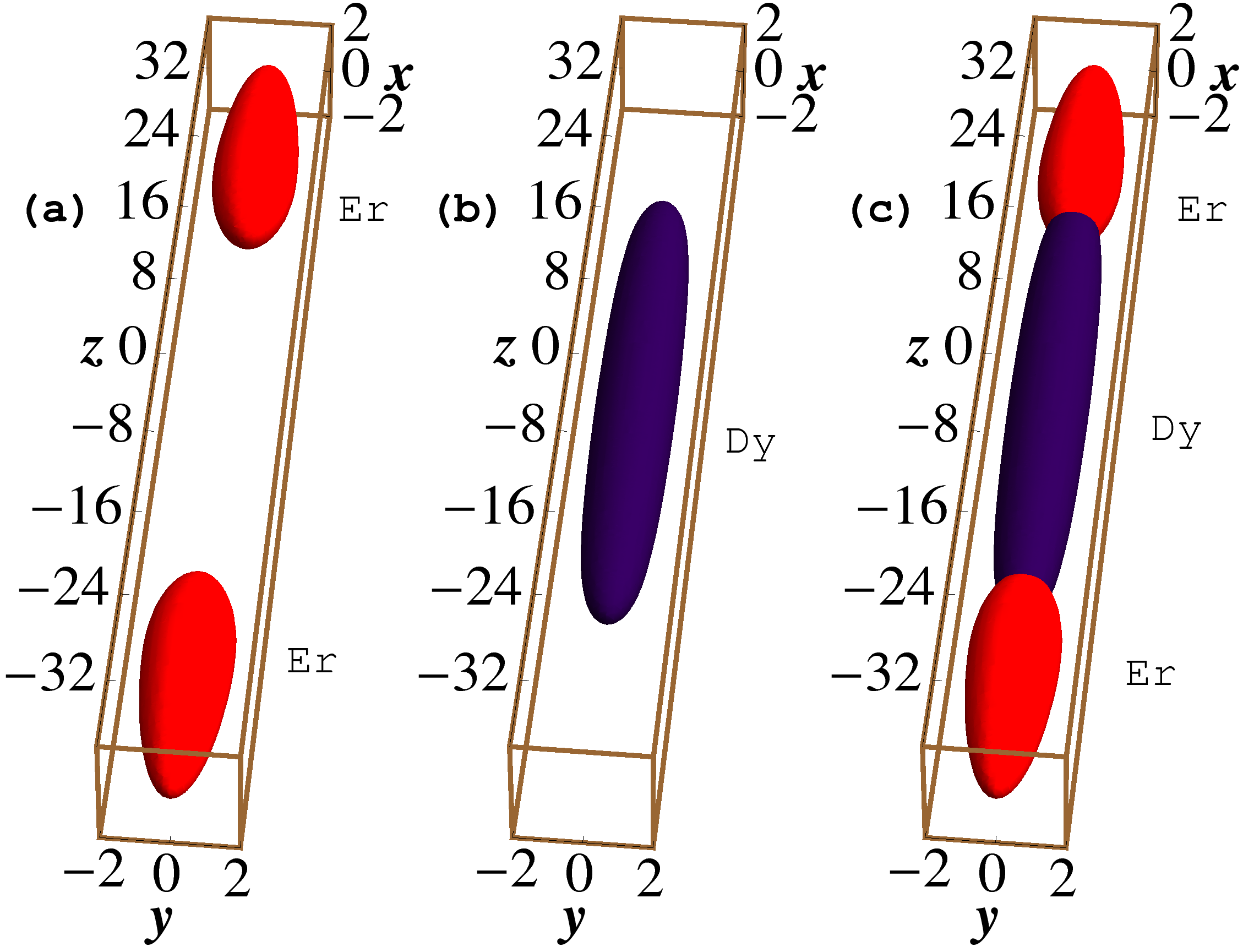}

\caption{ (Color online) 3D isodensity contour
(a) $|\phi_2|^2$  of $^{168}$Er, 
(b) $|\phi_1|^2$  of $^{164}$Dy, and (c) $(|\phi_1|^2+|\phi_2|^2)$ of
the $^{164}$Dy-$^{168}$Er mixture   for the
binary soliton of 5000 $^{164}$Dy and 
1000 $^{168}$Er atoms with  $a($Dy$)= 120a_0, a($Er$)= 60a_0, a($Dy-Er$)=90a_0$. The dimensionless lengths $x, y,$ and $z$ are in units
of $l_0(\equiv 1$ $\mu$m).
The density on contour is 0.002.
}\label{fig6} \end{center}

\end{figure}

In Fig. \ref{fig3} we presented a spatially-symmetric demixed binary dipolar soliton with 
 2000 $^{164}$Dy and 5000 $^{168}$Er atoms. In this case, $^{168}$Er atoms lie in the 
central region  whereas $^{164}$Dy atoms move away from the central region. The situation changes with a larger number of $^{164}$Dy atoms. This is illustrated Fig. \ref{fig6} with 5000 $^{164}$Dy atoms and 1000 $^{168}$Er atoms.  In this case, in the  isodensity contour of the binary soliton the roles of $^{164}$Dy and $^{168}$Er atoms are changed compared to that in Fig. \ref{fig3}. 
Now, contrary to that in Fig. \ref{fig3},  the $^{164}$Dy atoms occupy the central region and the $^{168}$Er atoms occupy the peripheral region. In Figs. \ref{fig4} and \ref{fig5} in the darker pink region in the spatially-symmetric configuration the $^{164}$Dy atoms stay at the center and $^{168}$Er atoms stay in the outer region and the opposite happens in the darker green region. 
 The binary  soliton of Fig. \ref{fig3}
lies in the darker green area  of Fig. \ref{fig4} (b), and  the binary  soliton of Fig. \ref{fig6}
lies in the darker pink area  of Fig. \ref{fig5} (b).

\begin{figure}[!t]

\begin{center}
\includegraphics[width=\linewidth]{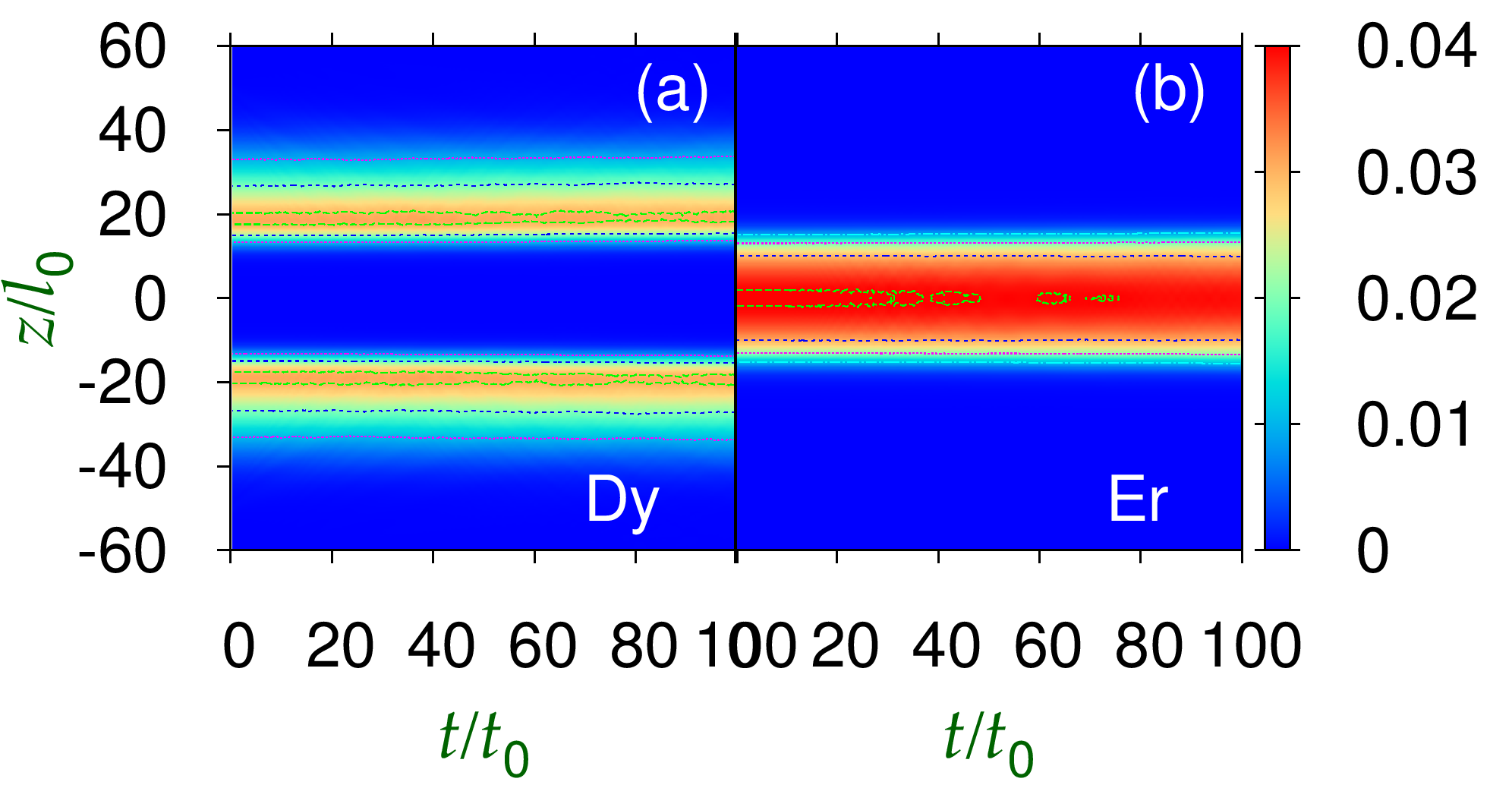}

\caption{ (Color online)  
Contour plot of integrated 1D densities $|\phi_i(z,t)|^2=\int dx\int dy |\phi({\bf r},t)|^2 $ of $^{164}$Dy and $^{168}$Er atoms when the 
binary soliton of Fig. \ref{fig3} is propagated with the time-dependent interspecies scattering length 
$a($Dy-Er$) = 90a_0+10a_0 \cos( t/t_0)$ with all other parameters    maintained at their time-independent initial values.
}\label{fig7} \end{center}

\end{figure}

\begin{figure}[!t]

\begin{center}

\includegraphics[width=\linewidth]{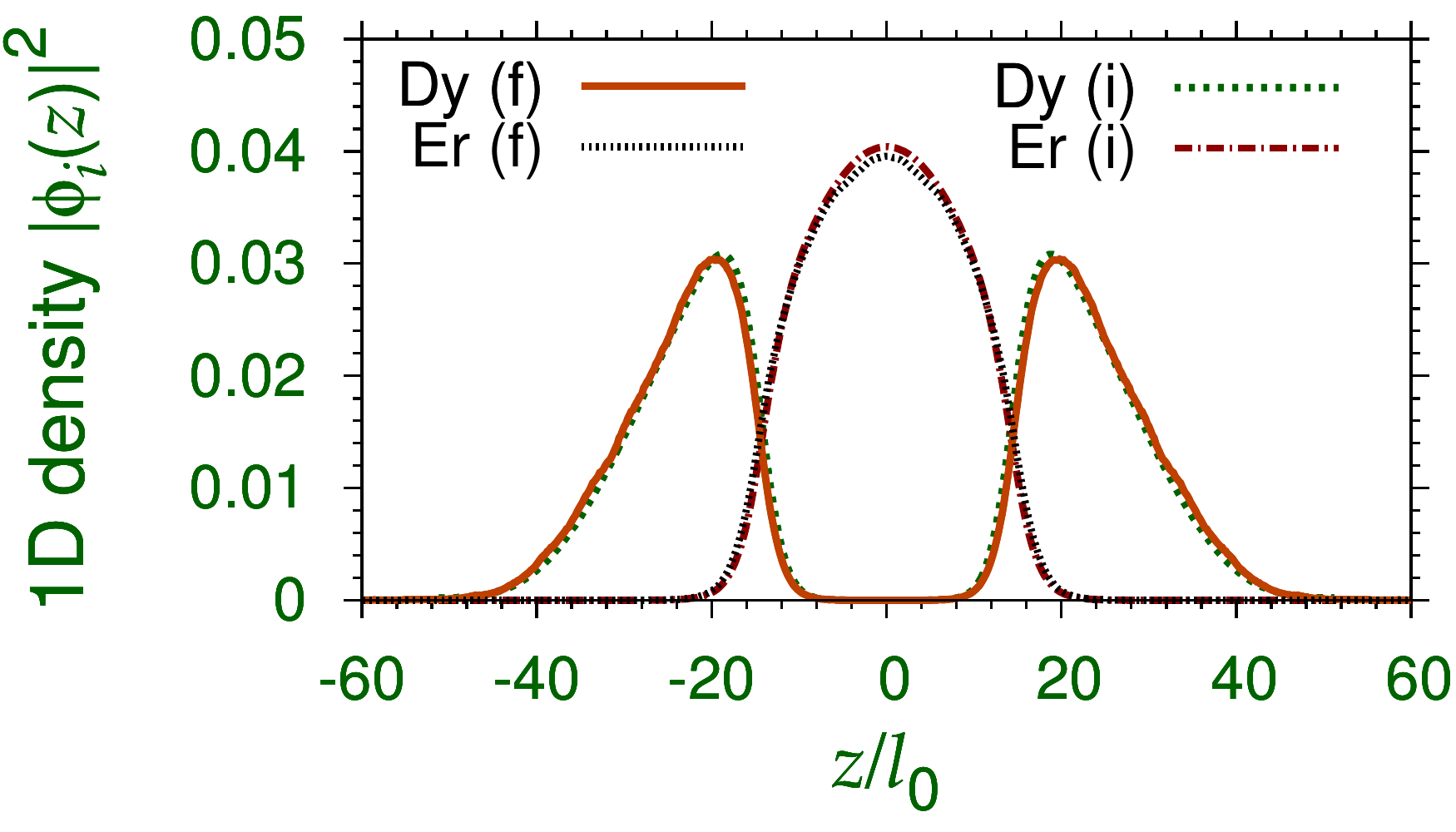}

\caption{ (Color online) Initial (i) at $t=0$ and final (f) at $t=100$ values of 
  integrated 1D densities $|\phi_i(z)|^2=\int dx\int dy |\phi({\bf r})|^2$ of $^{164}$Dy and $^{168}$Er 
atoms in the binary $^{164}$Dy-$^{168}$Er soliton of Fig. \ref{fig3} during real-time propagation of Fig. \ref{fig7}. 
}\label{fig8} \end{center}

\end{figure}

It is important to establish the stability of these spatially-symmetric solitons under small variation of the parameters of the initial state. To achieve this, we consider the final state of the spatially-symmetric binary soliton as obtained from the imaginary-time routine and use it as the initial state in the real-time program. First, we perform the numerical simulation with the binary soliton of Fig. \ref{fig3} with the time-dependent interspecies scattering length 
$a_{12}= 90a_0+10 a_0\cos (t/t_0)$, while the initial state was obtained with $a_{12}=90a_0$.
All other contact and dipolar interaction strengths are maintained at their initial values.     
This perturbation should be sufficient to destroy an unstable soliton.   The system executes small oscillation around the stable position.  Otherwise, no noticeable change is observed after 100 units of real-time propagation.   Our results are illustrated in Figs. \ref{fig7} and \ref{fig8}. In Fig.  \ref{fig7} we illustrate the contour plot of integrated 1D density
$|\phi_i(z,t)|^2 = \int  \int  dx   dy |\phi(x,y,z,t)|^2$ during real-time propagation. The visual profile of the solitons in Fig. \ref{fig7} maintain a constant width and 
do not show a sign of instability. To have a more quantitative view, in Fig. \ref{fig8} we plot the initial and final 1D densities $|\phi_i(z,t)|^2$ at times $t/t_0=0$ and 100 during  real-time propagation. This clearly demonstrates the stability of the binary dipolar soliton under small oscillation.

\begin{figure}[!t]

\begin{center}
\includegraphics[width=.8\linewidth]{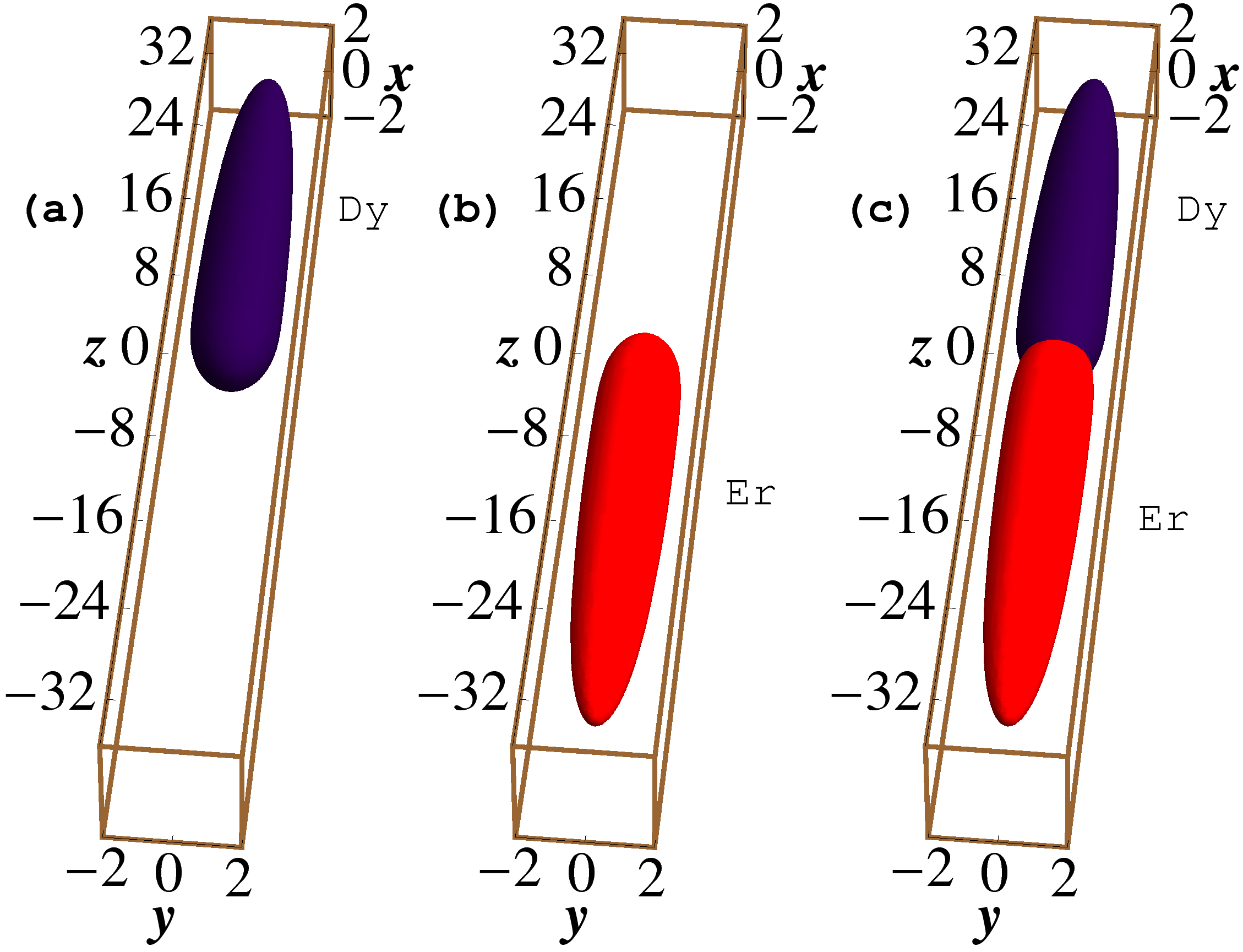}

\caption{ (Color online)  Same as in Fig. \ref{fig3} for the 
symmetry-broken 
binary  soliton of 2000 $^{164}$Dy and 
5000 $^{168}$Er atoms.  
}\label{fig10} \end{center}

\end{figure}

The   binary dipolar solitons studied above  were spatially-symmetric. One can also have symmetry-broken spatially-asymmetric binary solitons for the same parameters used as in the case of spatially-symmetric binary solitons. However, to generate these solitons by imaginary-time propagation one should consider spatially-asymmetric initial states.   
In particular we consider the following spatial-symmetry broken (asymmetric around $z=0$)
initial states with a spatial displacement  between the two profiles in imaginary-time propagation:
\begin{eqnarray}\label{x}
\phi_1({\bf r}) =\frac{\pi^{-3/4}}{w_\rho \sqrt w_z} \exp\left[  -\frac{\rho^2}{2w_\rho^2}     -\frac{(z-z_0)^2}{2w_z^2}   \right] \\
\phi_2({\bf r})= \frac{\pi^{-3/4}}{w_\rho \sqrt w_z} \exp\left[  -\frac{\rho^2}{2w_\rho^2}     -\frac{(z+z_0)^2}{2w_z^2}   \right]
\label{y}
\end{eqnarray}
in place of the initial states (\ref{eq5}).  It is important to mention that the imaginary-time propagation converges to the lowest-energy state  maintaining  the symmetry of the  initial state.
For example, in the 1D harmonic oscillator problem the imaginary-time propagation with a spatially-symmetric initial state will converge to the ground state of the problem, whereas with  
a spatially-antisymmetric initial state it will converge to the spatially-antisymmetric first excited state. With the initial states (\ref{x})  and (\ref{y}), we obtain the  symmetry-broken spatially-asymmetric binary solitons with the same parameters as used in the  spatially-symmetric binary solitons. We tried some other forms of symmetry-broken initial state and found that the 
solution converges always to the same final state. After some experimentation, we could locate for the same set of parameters of the problem only two definite demixed final states: (a) the symmetric states of Figs. \ref{fig3} and \ref{fig6}  and (b) the symmetry broken states to be presented next. 

  \begin{figure}[!t]

\begin{center}
\includegraphics[width=.8\linewidth]{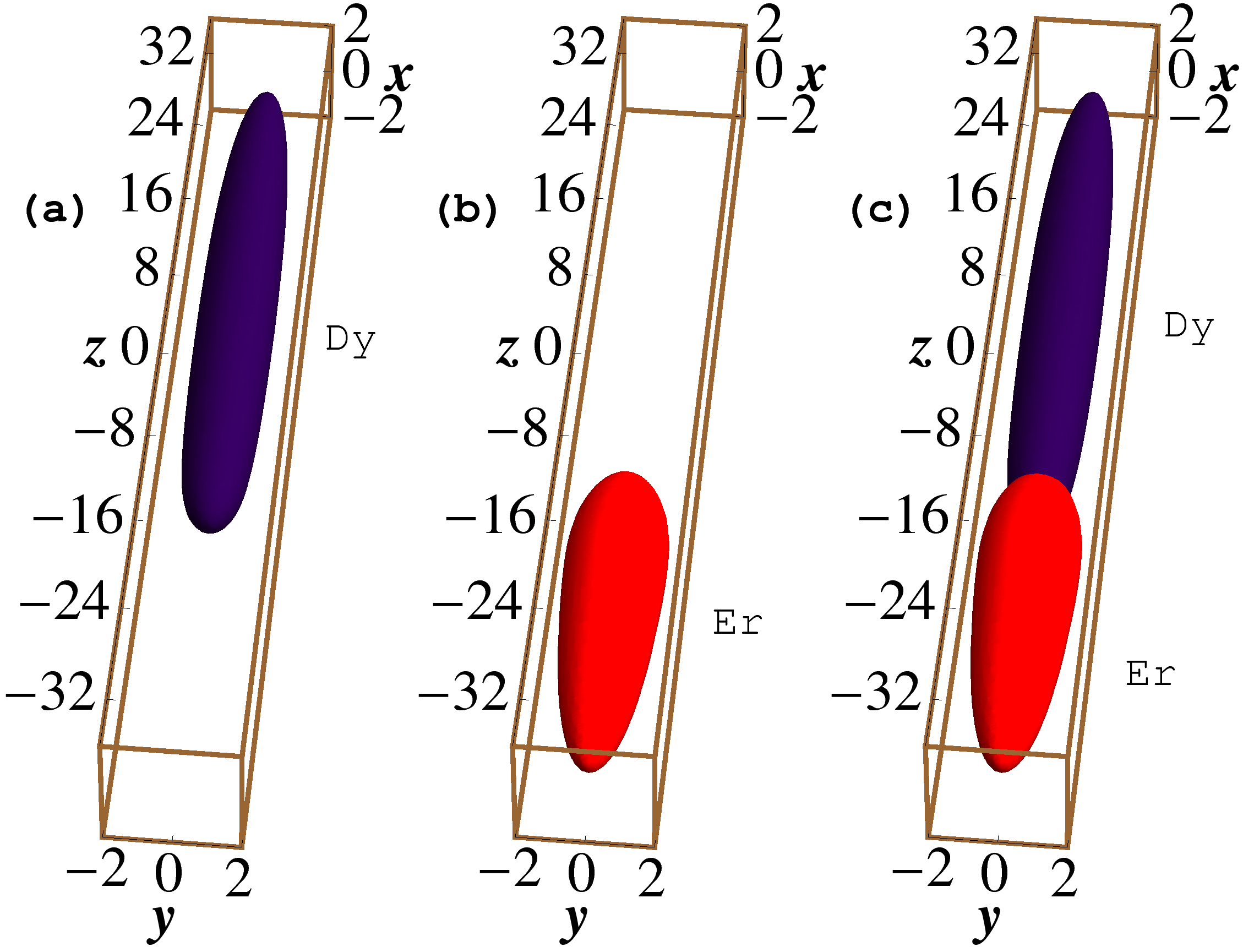}

\caption{ (Color online)  Same as in Fig. \ref{fig3} for the 
symmetry-broken 
binary  soliton of 5000 $^{164}$Dy and 
1000 $^{168}$Er atoms.  
}\label{fig11} \end{center}

\end{figure}

We perform  imaginary-time propagation with the initial states (\ref{x})  and (\ref{y}) using the same parameters as in Fig. \ref{fig3}. 
The result was insensitive to the value of the parameter $z_0$ in Eqs. (\ref{x}) and (\ref{y}) and we take $z_0=10$ in this study.   The 3D isodensity contours of the resultant soliton profiles in this case for 2000 $^{164}$Dy and 5000 $^{168}$Er atoms 
are shown in Fig. \ref{fig10}.  The converged binary dipolar soliton is clearly spatially-asymmetric. The second species with a larger number of $^{168}$Er atoms  has a longer spatial extension than the $^{164}$Dy atoms. Then we consider imaginary-time propagation 
using the same parameters as in Fig. \ref{fig6}. The 3D isodensity contours of the converged state are shown in Fig. \ref{fig11}.  In this case with 5000 $^{164}$Dy and 1000 $^{168}$Er atoms the $^{164}$Dy BEC clearly  occupies a larger region of space than the $^{168}$Er atoms. 

{ In case of demixed solitons, it is interesting to ask which of the two configurations $-$ symmetric or asymmetric $-$ has the lower energy and hence will be experimentally favorable.  To this end we calculated numerically the respective energies in several cases and found that consistently the spatially asymmetric configuration has slightly  
lower energy. However, this will  not be of phenomenological interest as the energy difference is very small, being  of the order of 0.1$\%$ of the total energy of either state.}

\begin{figure}

\begin{center}
\includegraphics[width=\linewidth]{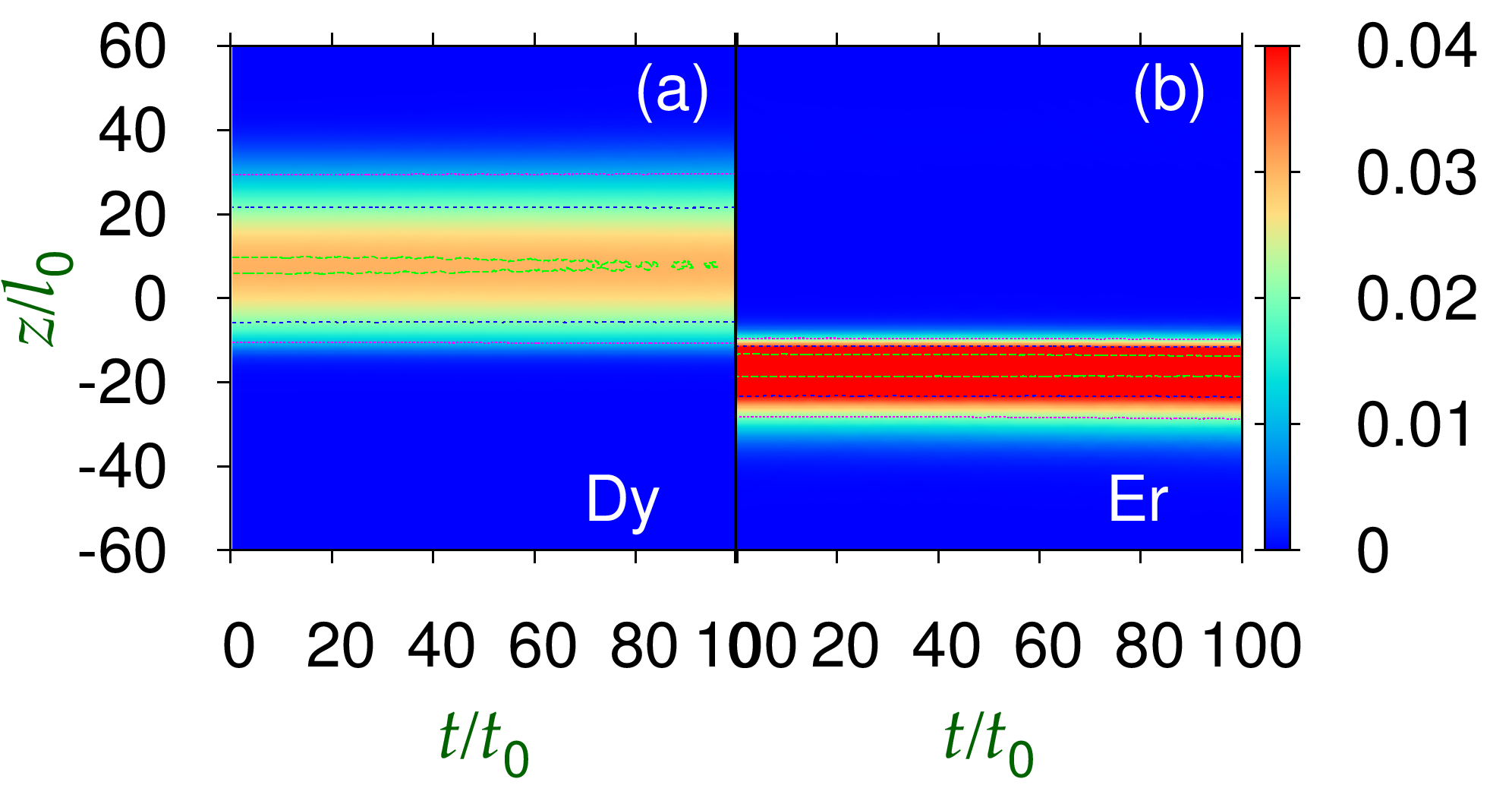}

\caption{ (Color online)  
Same as in Fig. \ref{fig7} for the symmetry-broken binary soliton of Fig. \ref{fig11}. 
%Contour plot of integrated 1D densities $|\phi_i(z,t)|^2$ of $^{164}$Dy and $^{168}$Er atoms when the initial 
%symmetry-broken binary soliton of Fig. \ref{fig11} is propagated with the time-dependent intraspecies scattering length 
%$a($Dy-Er$) = 90a_0+10a_0 \cos (t/t_0)$.   The other scattering lengths and all dipolar interaction strengths are maintained at their time-independent initial values during 
%time evolution.
}\label{fig12} \end{center}

\end{figure}

Now we investigate the stability of these symmetry-broken binary dipolar solitons. To this end we consider the converged solutions of the imaginary-time routine and use these as the initial states in real-time routine with the oscillating interspecies scattering length $a_{12}=a($Dy-Er$)
 = 90a_0 +10a_0\cos (t/t_0)$ as in the case of oscillation presented in Fig. \ref{fig7}. During real-time propagation the system executes small oscillation and the contour plot of the integrated 
1D densities is illustrated in Fig. \ref{fig12}. There is no visible change in the widths of the condensates during real-time propagation confirming stability. We also compared the 1D and 3D profiles 
of the initial and final states (not presented here) and found that they are practically identical.

The demixed binary solitons as presented in Figs. \ref{fig3}, \ref{fig6}, \ref{fig10}, e \ref{fig11} are ideal examples of soliton molecules of different species where the participating solitons maintain their identities with a minimum of overlap. The binding between the two solitons come from the long-range dipolar interaction. In nondipolar systems such binding can only come from short-range interspecies attraction and a composite soliton molecule can appear possibly only with complete overlap between participating solitons \cite{mfb3,solmol}.

\begin{figure}[!t]

\begin{center}
\includegraphics[width=\linewidth]{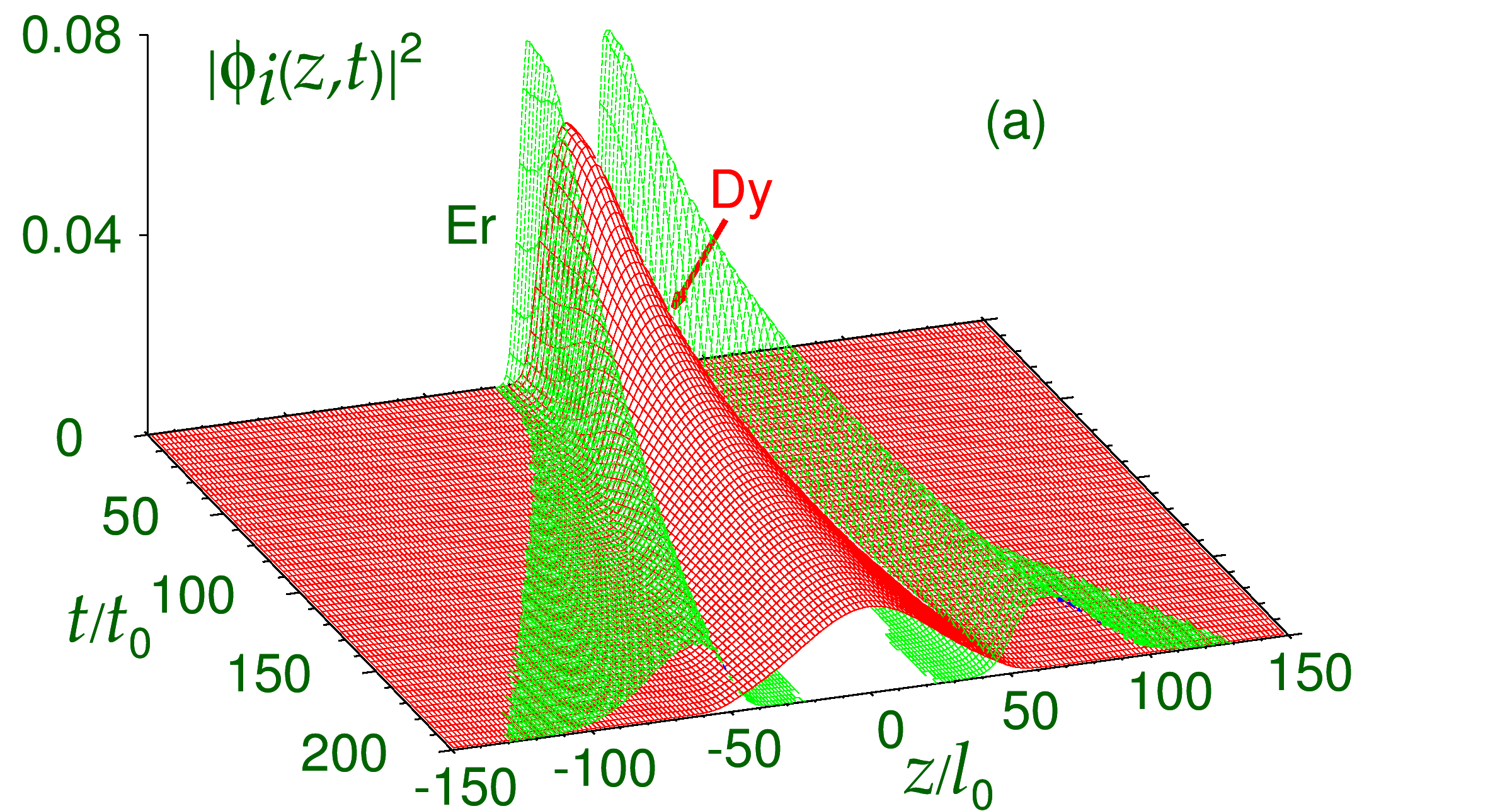}
\includegraphics[width=\linewidth]{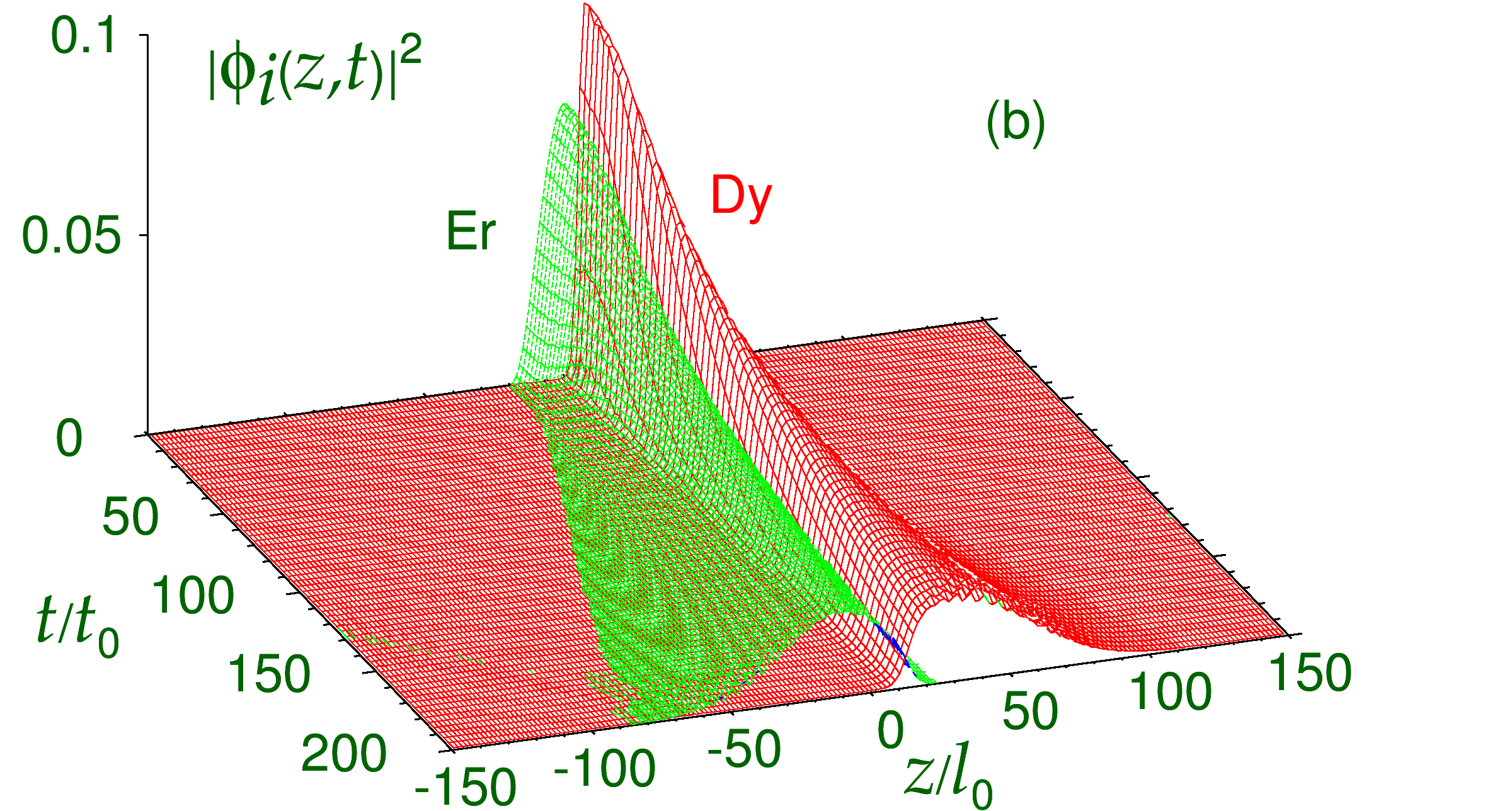}
\caption{ (Color online)
Integrated 1D density $|\phi_i(z,t)|^2=\int dx\int dy |\phi_i({\bf r},t)|^2 $ of $^{164}$Dy (red) and $^{168}$Er (green) atoms during real-time propagation
when the axial trap of angular frequency $\omega_z = 2 \pi \times 2$ Hz 
on a (a)  quasi-1D  spatially-symmetric binary dipolar BEC of 5000  $^{164}$Dy and 1000 $^{168}$Er 
atoms and a
(b) quasi-1D spatially-asymmetric binary  dipolar BEC of 2000 $^{164}$Dy and 5000 $^{168}$Er 
atoms 
is removed linearly for $20>t/t_0>0$ and the resultant 
soliton is propagated.  
Parameters used in simulation:  
$a($Dy$)=120a_0,a($Er$)=60a_0, a($Dy-Er$)= 90a_0,
 \omega_\rho=2\pi \times 61$ Hz, $l_0=1$ $\mu$m. 
}\label{fig13} \end{center}

\end{figure}

The demixed binary dipolar soliton or the binary soliton molecule can be prepared and observed experimentally. We illustrate  the possibility of  experimental realization of the soliton shown in 
Figs. \ref{fig6} and   \ref{fig10} by realistic numerical simulation using the binary GP Eqs. (\ref{eq3}) and (\ref{eq4}).  
In case of the spatially-symmetric binary soliton of Fig. \ref{fig6}
 we prepare a spatially-symmetric binary quasi-1D trapped dipolar BEC of 5000 $^{164}$Dy and 1000 $^{168}$Er atoms with the same parameters as in Fig. \ref{fig6} 
by imaginary-time simulation
but with traps $V({\bf r})=(\rho^2/2+\lambda^2z^2/2)$ and $V({\bf r})=m_\omega(\rho^2/2+\lambda^2z^2/2)$ with $\lambda =0.03$, respectively, on $^{164}$Dy and $^{168}$Er atoms, in place of the axially free traps  $V({\bf r})=\rho^2/2$ and $V({\bf r})=m_\omega\rho^2/2$. 
For both  $^{164}$Dy and $^{168}$Er atoms these traps correspond to one of axial angular frequency $\omega_z=2\pi \times 
2$ Hz and of radial angular frequency $\omega_\rho=2\pi\times 61$ Hz.
This binary bound dipolar BEC is then used 
as the initial state in the real-time program.   During real-time propagation, from $t/t_0=0$ to 20  
the axial traps $\lambda^2 z^2/2$ are gradually (linearly) reduced to zero, so that for $t/t_0>20$ 
the axially free quasi-1D binary dipolar soliton emerges. The simulation is continued for a long interval of time and a steady propagation of the binary demixed dipolar soliton is established. 
The outcome of this simulation is shown in Fig. \ref{fig13} (a),    where we plot the integrated 
1D density $|\phi_i(z,t)|^2$ of the two component solitons as obtained from real-time propagation.  A similar simulation for the spatially-asymmetric binary soliton of Fig. \ref{fig10} is shown in Fig. \ref{fig13} (b).  After being released from the axial trap for $t/t_0>20$,
the solitons expand a little and then oscillate around their appropriate sizes.   
The stability of the demixed binary dipolar solitons is evident in the simulations illustrated in Figs. \ref{fig13} (a) and (b).

\begin{figure}[!t]

\begin{center}
\includegraphics[width=\linewidth]{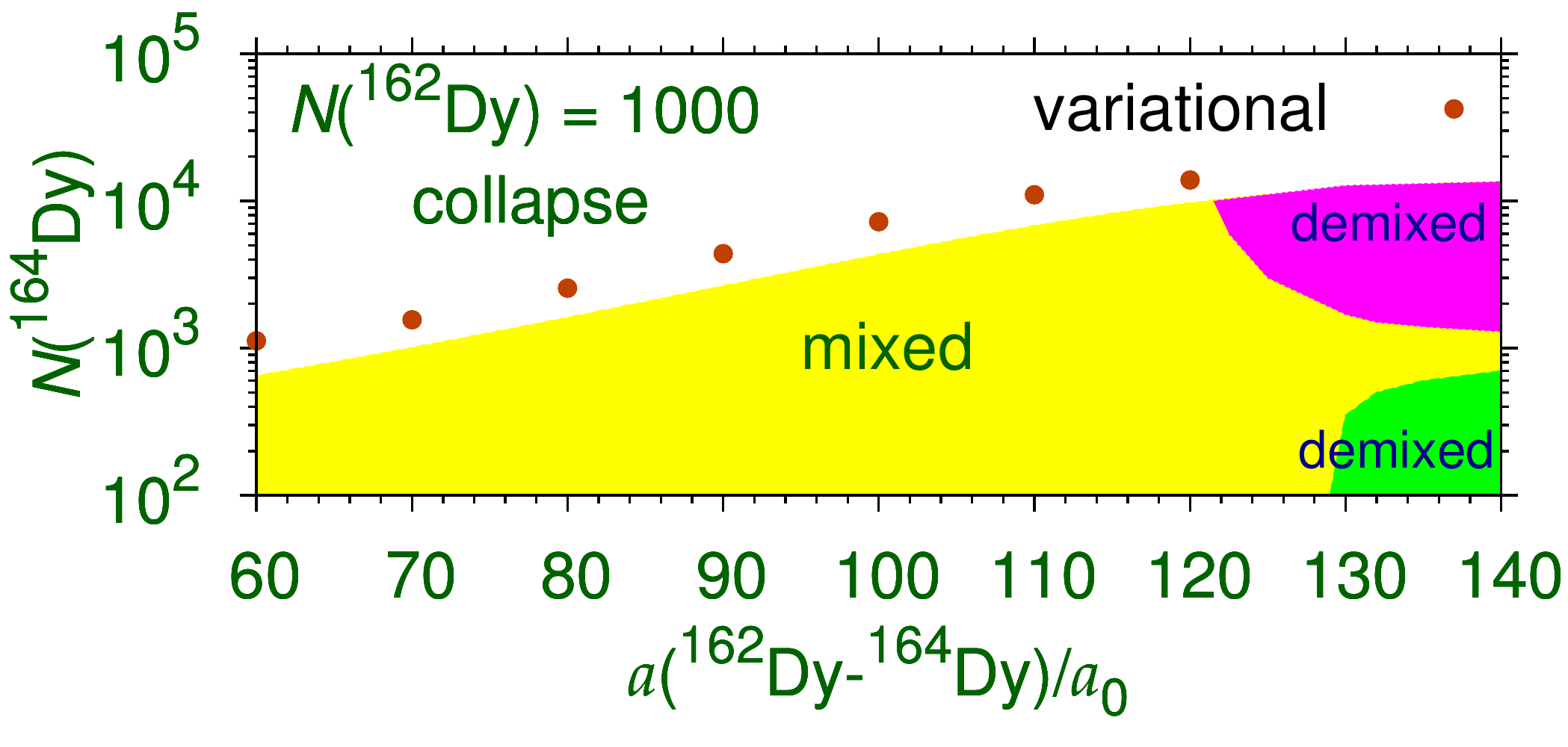}

\caption{ (Color online) Same as in Fig. \ref{fig4} for a binary 
$^{164}$Dy-$^{162}$Dy soliton  for  $N$($^{162}$Dy) = 1000 atoms.
The variational mixed-collapse boundary is indicated by solid circles. 
  For a spatially-symmetric demixed binary soliton,
in the upper darker (pink) region the  $^{162}$Dy atoms stay out and in the lower darker (green)
region the $^{164}$Dy atoms stay out of the central region. 
Parameters used: $l_0 = 1$  $\mu$m, $a(^{164}$Dy)= $a(^{164}$Dy)=
120$a_0$.
}\label{fig14} \end{center}

\end{figure}

\subsection{Binary $^{164}$Dy-$^{162}$Dy soliton}

Here we consider yet another type of binary dipolar soliton, e.g. the $^{164}$Dy-$^{162}$Dy soliton. This is particularly interesting as Lev and his collaborators are studying this binary mixture in laboratory \cite{levb}. In order to permit a large number of atoms in the solitons we consider a large value for the scattering lengths, e.g.
$a(^{162}$Dy) = $a(^{164}$Dy) =120$a_0$, viz. Fig. \ref{fig1}. The interspecies scattering length is considered as a variable.  
First we consider the stability phase plot for this binary soliton. In Fig. \ref{fig14} we show the number of $^{164}$Dy atoms in the binary soliton for 1000 $^{162}$Dy atoms. Again there appears mixed and demixed binary solitons.  
As the mass and dipolar lengths are almost the same for the two isotopes the binary plot is quasi symmetric.  The plot for   1000 $^{164}$Dy atoms in the binary soliton will lead to  a practiaclly identical plot as that in Fig. \ref{fig14} with the role of the two isotopes interchanged.  This system is highly dipolar compared to the 
$^{164}$Dy-$^{168}$Er mixture and hence  can accommodate a smaller number of solitons as can be seen by comparing Fig. \ref{fig14} with Figs. \ref{fig4} and \ref{fig5} with 1000 atoms of one of the components. Compared to the 
$^{164}$Dy-$^{168}$Er mixture,  demixed binary solitons in 
$^{164}$Dy-$^{162}$Dy only appears for a larger value of interspecies scattering length to compensate for larger interspecies dipolar attraction. In Figs. \ref{fig4} and \ref{fig5} demixing appears for 
interspecies scattering length $a_{12}\lesssim 100a_0$, whereas in Fig. \ref{fig14} demixing appears for $a_{12}\gtrsim 120a_0$.   

\begin{figure}[!t]

\begin{center}
\includegraphics[width=.8\linewidth]{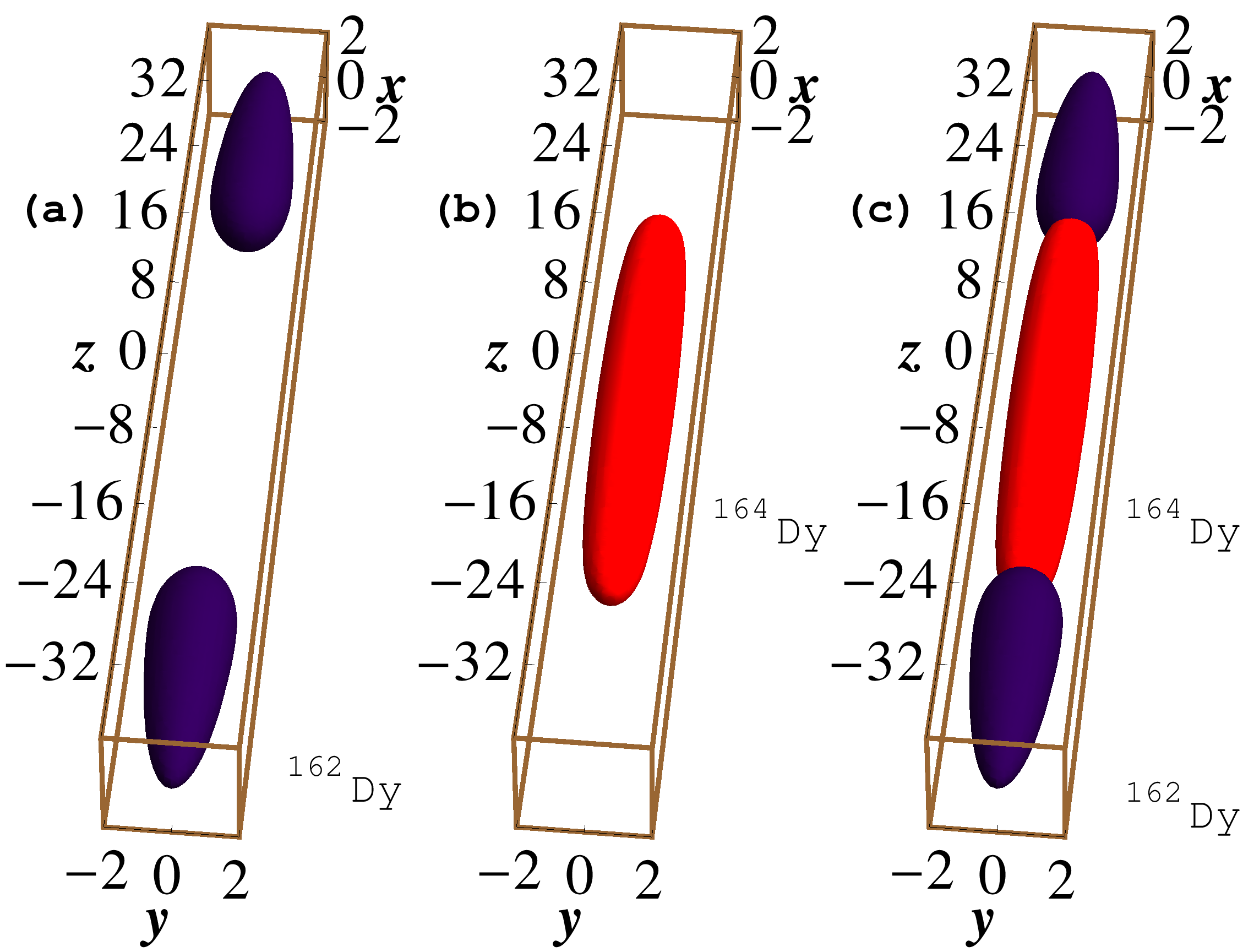}

\caption{ (Color online) 3D isodensity contour 
(a) $|\phi_1|^2$ of   $^{162}$Dy, 
(b) $|\phi_2|^2$  of $^{164}$Dy, and (c) $(|\phi_1|^2+|\phi_2|^2)$ of 
the $^{164}$Dy-$^{162}$Dy mixture   for the binary soliton of 1000 $^{162}$Dy and 
5000 $^{164}$Dy atoms with  $a(^{162}$Dy$)= a(^{164}$Dy$)= 120a_0, 
 a(^{162}$Dy-$^{164}$Dy)=$130a_0$. 
The dimensionless lengths $x, y$ and $z$ are in units
of $l_0(\equiv 1$ $\mu$m).
The density on contour is 0.002.
}\label{fig15} \end{center}

\end{figure}

In this case also one has mixed, spatially-symmetric demixed, and spatial symmetry-broken demixed binary solitons. 
Without repeating a detailed discussion of different types of solitons, 
in Fig. \ref{fig15} we show the isodensity contour of a typical 
spatially-symmetric
binary $^{164}$Dy-$^{162}$Dy soliton for 1000  $^{162}$Dy atoms and 5000 
$^{164}$Dy atoms for the interspecies scattering length 
 $a(^{162}$Dy-$^{164}$Dy)=$130a_0$. The component $^{162}$Dy  with a 
smaller number of atoms stays out of the central region whereas the component  $^{164}$Dy with a larger number of atoms stays at the center.

\section{Summary and Discussion} 

Using  the numerical solution of a set of coupled 3D mean-field GP equations, 
we demonstrate the existence
 of demixed    dipolar binary  $^{164}$Dy-$^{168}$Er 
and $^{164}$Dy-$^{162}$Dy  solitons stabilized  by
inter- and intraspecies  
dipolar interactions in the presence of repulsive inter- and intraspecies contact interactions.
  The domain of the appearance  of the binary soliton 
 is highlighted in stability diagrams of number of atoms in the two components
and interspecies scattering length
$a_{12}$ for fixed dipolar  and intraspecies contact interactions.
The binary soliton is stable for a maximum number of atoms beyond which it collapses. For small interspecies interaction $a_{12}$ the binary soliton is mixed and for large $a_{12}$ it is demixed. 
The mixed-collapse boundary was also calculated using a Gaussian variational approximation \cite{mfb3} and was found to be in reasonable agreement with the numerical result. 
We found two types of demixing in this case: (a) spatially-symmetric and 
(b) spatial-symmetry-broken  demixed binary solitons. In the spatially-symmetric case the 
species with smaller number of atoms breaks up into two equal parts and the  parts leave the central region occupied by the species with larger number of atoms and consolidate on both sides of the BEC component with larger number of atoms as shown in Figs. \ref{fig3},
\ref{fig6} and \ref{fig15}. 
In the spatially-asymmetric case,  the two solitons move away from each other and finally stabilize side by side as shown in Figs. \ref{fig10} and \ref{fig11}. 
The two species of the demixed soliton have a minimum of overlap and are stabilized by dipolar interaction in the presence of reasonably strong contact interaction. Such demixed solitons are not possible in the absence of dipolar interaction. 
We also tested the stability of the demixed solitons by considering real-time propagation with
oscillating scattering length with small amplitude. A stable binary soliton should then execute 
small oscillations. The converged solution of the binary soliton in imaginary-time propagation 
is used as the initial state of real-time propagation.  The solitons are found to exhibit small oscillation over a long time thus establishing their stability.

The solitons considered in this paper are stabilized by long-range dipolar 
attraction and short-range contact repulsion. The dipolar interaction is attractive in the axial 
direction and the dipolar repulsion in the transverse direction is 
  compensated  by a harmonic  trap. Hence unlike normal BEC 
solitons stabilized by short-range contact attraction, the present dipolar 
BECs will be more immune to collapse due to short-range repulsion
and can easily accommodate 10000 atoms of the  binary $^{164}$Dy-$^{168}$Er mixture as can be
seen from Figs. \ref{fig4}, \ref{fig5}  and  \ref{fig14}.  The dipolar  ($N a_{\mathrm {dd}}$) and contact ($Na$) nonlinear strengths for such these 
solitons could be typically $\sim 50$ for 5000 $^{164}$Dy atoms. 
A nondipolar soliton is created only by contact attraction without any repulsion 
and is fragile due to collapse instability. In a stable nondipolar soliton, 
the maximum contact interaction strength is $N|a|\sim 0.6$ \cite{1,3,4}. 
Thus the dipolar soliton  with larger interaction strengths can accommodate a much larger number of atoms than the
nondipolar solitons and should 
be of great experimental interest.
A way of experimentally realizing these demixed dipolar solitons is suggested and the viability of this scheme is demonstrated by real-time simulation.  First, a quasi-1D
demixed binary BEC is to be formed with a weak trap in the axial $z$ direction. Then the axial  trap is to be removed linearly 
in a reasonably short time, while the  demixed binary BEC turns into a demixed binary soliton. 
 With the present experimental techniques, 
such binary dipolar BECs can be observed and the conclusions of this study verified.

\acknowledgments
We thank  
FAPESP  and  CNPq (Brazil)  for partial support.

\end{document}